\documentclass[12pt]{iopart}
\usepackage{framed,xcolor}
\usepackage{morefloats}
\usepackage{txfonts}
\usepackage{url}
\usepackage{graphicx}
\usepackage{threeparttable}
\usepackage{bm}
\usepackage{arydshln}
\usepackage{harvard}

\begin{document}

\renewcommand{\v}[1]
{   
\mathbf{#1}  
}
\newcommand{\m}[1]
{
 \mathbf{#1} 
}

\title[EEG Forward Modeling via H(div) Finite Element Sources with Focal Interpolation]{Electroencephalography (EEG) Forward Modeling via H(div) Finite Element Sources with Focal Interpolation}
\author{S Pursiainen$^{1}$ and J Vorwerk$^2$ and C H Wolters$^2$}
\address{$^1$Tampere University of Technology, Department of Mathematics, PO Box 553, FI-33101 Tampere, Finland} 
\address{$^2$Institute for Biomagnetism and Biosignalanalysis, University of M{\"u}nster, Germany}
\ead{sampsa.pursiainen@tut.fi}
\pagenumbering{arabic}

\begin{abstract}
The goal of this study is to develop focal, accurate and robust finite element method (FEM) based approaches which can predict the electric potential  on the surface of the computational domain given its structure and internal primary source current distribution. While conducting an EEG evaluation, the  placement of source currents to the geometrically complex grey matter  compartment is a challenging but necessary task to avoid forward errors attributable to tissue conductivity jumps.  Here, this task is approached via a mathematically rigorous formulation, in which the current field is modeled via divergence conforming H(div) basis functions. Both linear and quadratic functions are used while the potential field is discretized via the standard linear Lagrangian (nodal) basis. The resulting model includes dipolar sources which are interpolated into a random set of positions and orientations utilizing two alternative approaches: the position based optimization (PBO) and the mean position/orientation (MPO) method. These results demonstrate that the present dipolar approach can reach or even surpass, at least in some respects,  the accuracy of two classical reference methods, the partial integration (PI) and St.\ Venant (SV) approach which utilize  monopolar loads instead of dipolar currents. 
\end{abstract}

\section{Introduction}

\begin{figure}[h]
\begin{center} 
\begin{minipage}{5cm}
\begin{center}
    \includegraphics[width=3.2cm]{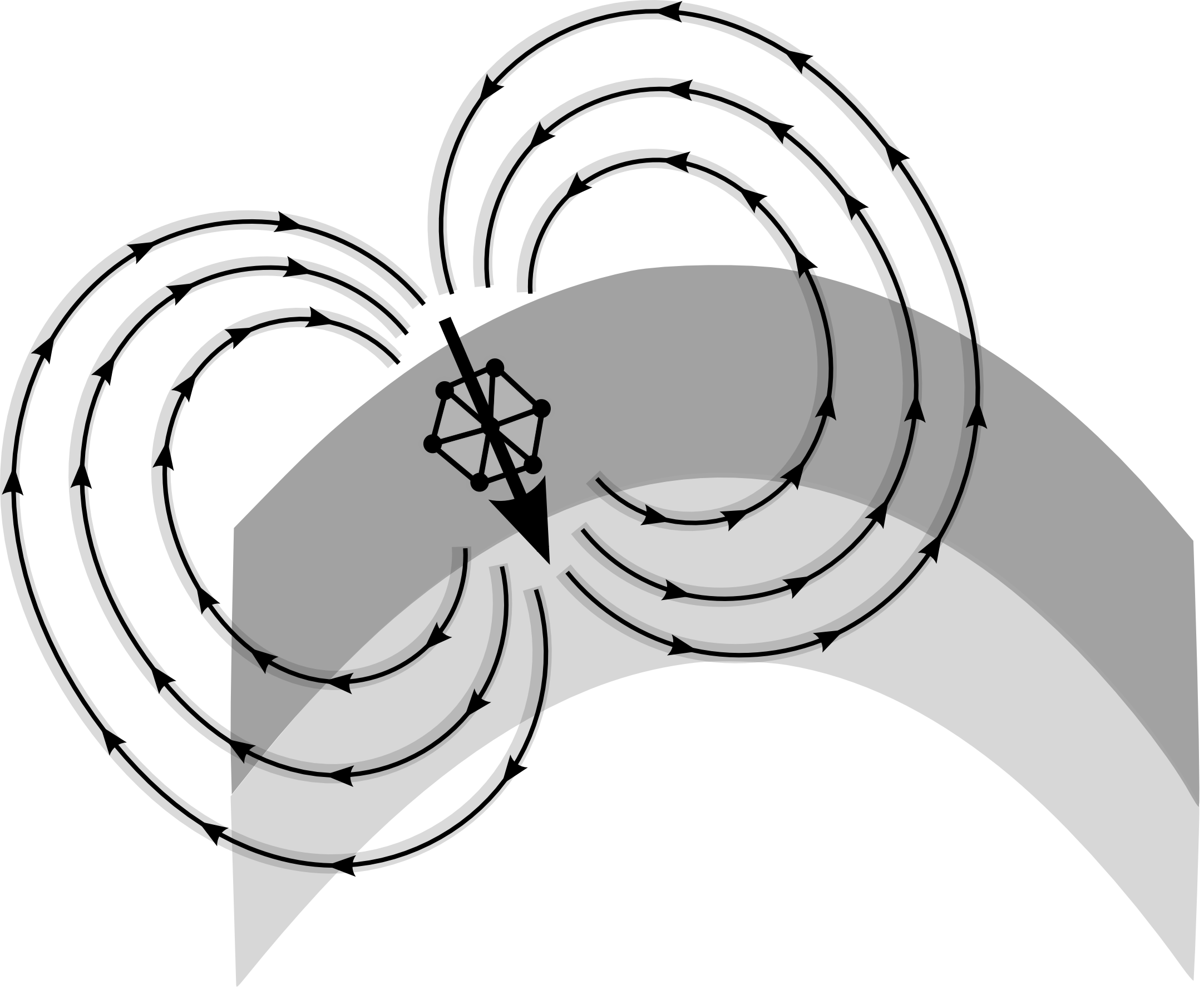} \\ Grey matter configuration
\end{center}
\end{minipage}  
\begin{minipage}{5cm}
\begin{center}
\includegraphics[width=3.2cm]{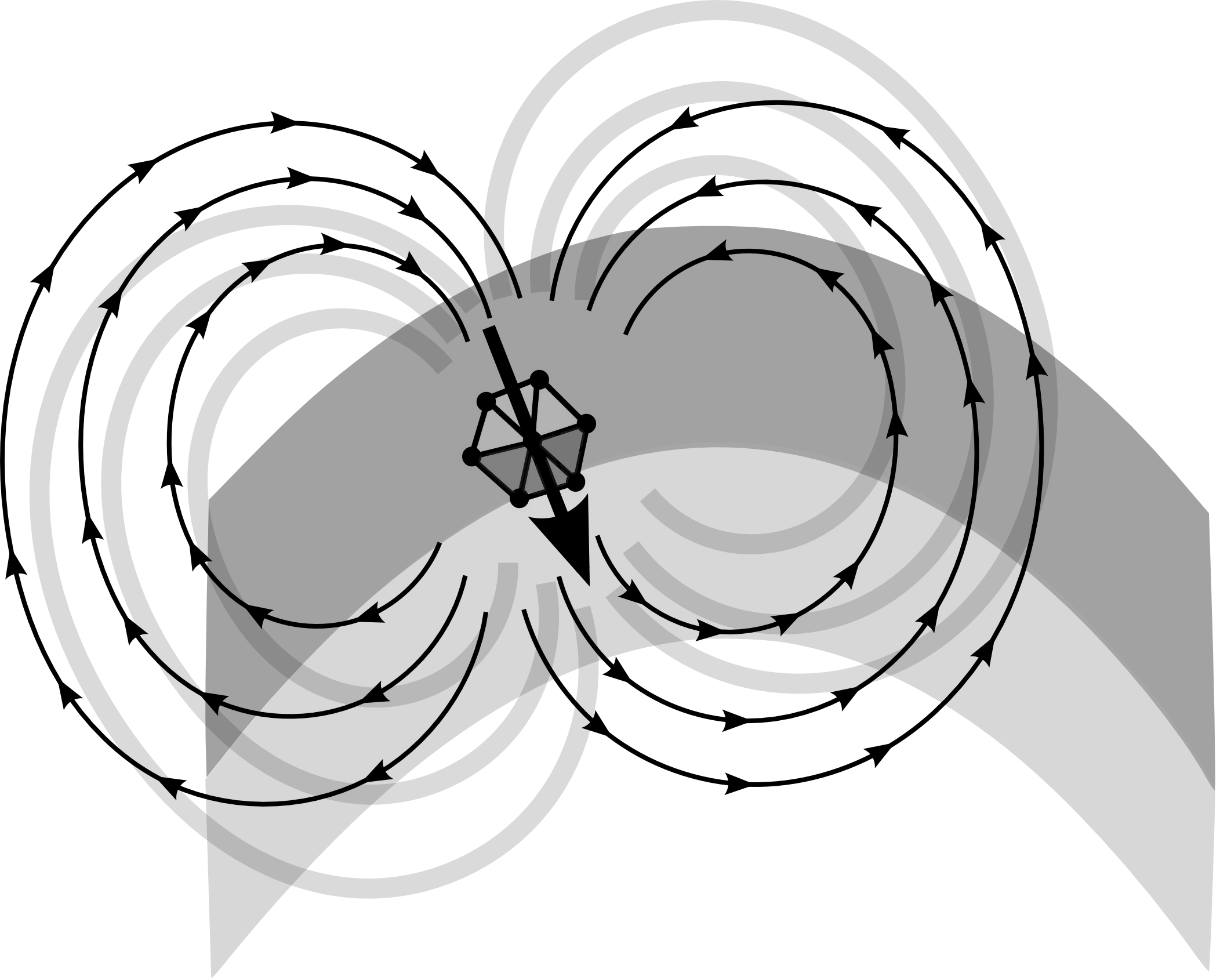} \\ Overlapping configuration
\end{center}
\end{minipage} 
\end{center}
    \caption{A schematic visualization of a situation in which a focal primary source current (large arrow) is approximated using a configuration of six finite element mesh nodes (point stencil). Left: If all the mesh nodes are located in the compartment of the grey matter (grey layer), then the approximation of the volume current distribution (countours with arrows), i.e., the conductivity multiplied by the negative gradient of the electric potential, matches with the actual one (light grey contours).  Right: If one or more nodes in the configuration overlap with compartment (light grey layer) other than that of the grey matter, then the volume current and thereby also the electric potential obtained will differ from the actual distribution due to the conductivity jump between the compartments. Consequently, the focality (compactness) of the source configuration is vital in order to avoid forward errors.   \label{contours}}
\end{figure}

The goal of this study is to develop focal (locally supported) approaches which can be incorporated into a finite element method (FEM) based bioelectromagnetic forward (data) simulation. We focus in particular on the electroencephalography (EEG) imaging of the brain activity   \cite{niedermeyer2004,brazier1961,hamalainen1993,CHW:Mun2012}. EEG is an electrophysiological measurement method, which monitors the electric potential distribution on the subject's head through a set of electrodes attached to the skin. Based on the measurements, the objective is to recover the primary source current distribution, that is, the neural activity of the brain. The forward problem is to calculate the electrode voltages given a fixed source current,  the geometry and  internal conductivity distribution of the head. The focal placement of the source currents to the geometrically complex grey matter compartment is a challenging but necessary task. Namely, if the simulated source is not focally restricted in the grey matter like the actual one, then it will result in an erroneous volume current and electric potential estimate (Figure \ref{contours}). An important aspect of the present source model is also the possibility to incorporate  {\em a priori} information about the tissue structure into the forward simulation, e.g., that of pyramidal neurons oriented normally to the grey matter compartment \cite{CHW:Creu62,CHW:Sch90}. This is especially interesting for the development of anatomically accurate forward models based on medical images of the brain  \cite{acar2010,CHW:Vor2014,aydin2015,fiederer2016}.  As the list of references shows, the authors are well aware of the state-of-the-art anatomical forward models. They have contributed to the development of very modern and realistic forward approaches for more than a decade now (see especially the overview book-chapter \cite{CHW:Mun2012}).   

The FEM is a flexible simulation tool for finding approximate solutions to boundary value problems for partial differential equations and, besides many other applications, also bioelectromagnetic problems \cite{braess2001,CHW:Mun2012}, especially, as it makes it possible to design and optimize the volumic tetrahedral mesh following precisely a given head geometry, including its internal folded surfaces \cite{CHW:Vor2014} and the three-dimensional conductivity structure. The other classical forward modeling approach, i.e., the boundary element method (BEM) \cite{CHW:Mun2012,CHW:Kyb2005,ataseven2008,CHW:Ste2012,ermer2001}, assumes that there is a layer-wise constant conductivity, which does not take into account detailed 3D structures, such as skull compacta and spongiosa \cite{CHW:Vor2014} or the anisotropic conductivity of the white matter \cite{Hal2008b,CHW:Gue2010,CHW:Vor2014}.  For a detailed introduction and overview of FEM-based EEG forward modeling techniques
we refer to \cite{CHW:Mun2012}. It becomes clear that, following from the
theory of the partial differential equations, finding the FEM solution of the EEG forward problem
necessitates the divergence of the source current to be square integrable, thereby ruling
out the classical (singular) dipole source.  Consequently, one has to rely on either subtraction methods  that are computationally very expensive and less accurate for high source eccentricities or on finitely supported primary source units which enable approximation of a given singular dipole as a limit, if the finite element (FE) mesh size tends to zero  \cite{CHW:Dre2009,CHW:Ber91,CHW:Mar98,CHW:Awa97,CHW:Sch2002}.  

In this paper, the forward model is tackled via a mathematically rigorous approach, in which the current field is modeled utilizing divergence conforming vector basis functions   \cite{ainsworth2003,monk2003,braess2001}. Both linear and quadratic functions are used while the potential field is discretized via the standard linear Lagrangian (nodal) basis. For a tetrahedral FE mesh, the resulting model can be associated with  dipolar face intersecting (FI)  and edgewise (EW)  sources. Of these, the FI sources correspond to the linear Whitney  (Raviart-Thomas) model \cite{pursiainen2011,pursiainen2012b,bauer2015} that is here supplemented with the EW orientations yielded by the quadratic extension.  

As a continuation of  a recent study \cite{bauer2015}, the FI and EW sources are explored both in non-interpolated and interpolated contexts. The following two alternative  interpolation approaches were investigated: the position based optimization (PBO) \cite{bauer2015} and the mean position/orientation  (MPO) method, which utilize a very compact set of four to eight nodal degrees of freedom shared by elements adjacent to the one containing the given position. 
The accuracy and robustness of these methods is explored in the numerical experiments for different focal vector source counts and configurations covering the support of maximally eight nodal basis functions. Two classical dipole estimation methods are used here as the reference: the partial integration (PI)  \cite{CHW:Yan91,weinstein2000,CHW:Awa97,CHW:Sch2002} and St.\ Venant approach  \cite{CHW:Sch94,buchner1997,toupin1965,medani2015}.  

The results obtained suggest that the present dipolar approach  can reach or even surpass, at least in some respects,  the accuracy of these reference methods. Hybrid source configurations including both FI and EW type orientations led to the best outcome with regard to both accuracy and numerical stability. 
Of the MPO and PBO interpolation methods, the first one yielded better results with respect to the overall accuracy and the second one with respect to stability and adaptability. 

This paper is organized as follows. Section \ref{materials_and_methods} gives a brief theory review of the forward model, dipolar sources and interpolation techniques. It also describes the numerical experiments. Section \ref{results} includes the results and Section \ref{discussion} consists of the discussion. Mathematical details of FI and EW sources can be found in the appendix (Section \ref{appendix}). 

\section{Materials and Methods}
\label{materials_and_methods}

\subsection{Forward Model}
\label{forward_model}

The present bioelectromagnetic forward problem is to predict the electric potential $u$ in  the closed domain $\Omega$, that is a volumetric model of a head,  given the symmetric and positive definite conductivity tensor $\sigma$ and the primary current field $\vec{J}^P$ in $\Omega$. The law of the  total charge conservation $\nabla \cdot \vec{J} = 0$  and the quasi-static  approximation of the  electromagnetism $\vec{J} = \vec{J}^P - \sigma \nabla u$ together yield the equation \cite{sarvas1987,hamalainen1993} \begin{equation} \nabla \cdot (\sigma \nabla u) = \nabla \cdot \vec{J}^P \quad \hbox{in} \quad \Omega \end{equation}  equipped with the boundary condition $(\sigma \nabla u) \cdot \vec{n} = 0$ on $\partial \Omega$ with $\vec{n}$ denoting the outward pointing normal vector. When multiplied by a test function $v$ and integrated by parts, this yields the weak form  
\begin{equation}
 \int_\Omega \nabla v \cdot ( \sigma \nabla u) \; dV = -\int_\Omega v (\nabla \cdot \vec{J}^P) \; dV
   \hspace{0.2cm} \hbox{for all} \; v \in H^1(\Omega),
  \label{weak}
\end{equation}
where the Sobolev space $H^1(\Omega)$ consists of functions with all first-order partial derivatives  square integrable, i.e., in $L_2(\Omega)$. Equation (\ref{weak}) has the solution $u \in  H^1(\Omega)$, unique up to choosing the zero level of the potential, if the divergence of the primary current density is square integrable, i.e., if $J^P \! \in \! H(\hbox{div}) \! = \! \{ \vec{w} \, | \, \nabla \cdot \vec{w} \! \in \! L^2(\Omega) \}$ \cite{evans1998,braess2001}.


The potential and primary current density are approximated via $u_h = \sum_{i=1}^N z_i \psi_i$ and $\vec{J}_h^P = \sum_{j = 1}^K x_j \vec{w}_j$, respectively,  where $\psi_1,\psi_2,\ldots,\psi_N$ are linear nodal basis functions belonging to $H^1(\Omega)$ and $\vec{w}_1, \vec{w}_2,\ldots,\vec{w}_K \in H(\hbox{div})$. The relation between the corresponding coordinate vectors, ${\bf z}= (z_1, z_2, \ldots, z_N)$ and ${\bf x}= (x_1, x_2, \ldots, x_K)$, is determined by the linear system \begin{equation} \label{jees} {\bf A} {\bf z}  = {\bf G} {\bf x}  \end{equation}  with ${\bf A} \in \mathbb{R}^{N \times N}$, ${\bf G} \in \mathbb{R}^{N \times K}$,   $A_{i,j}= \int_{\Omega} \nabla \psi_j \cdot (\sigma \nabla \psi_i) \; dV$, and  $G_{i,j}  = \int_\Omega  \psi_i (\nabla \cdot {\vec w}_j )  d V$. Given {\bf x} one can obtain {\bf z} by solving the system ${\bf A} {\bf z} = {\bf f}$ with load vector ${\bf f} = {\bf G} {\bf x}$ and, consequently, an electrode voltage vector ${\bf y} = {\bf R} {\bf z} $ can be formed as \begin{equation}  
{\bf y} =  {\bf R} {\bf A}^{-1} {\bf G} {\bf x} =  {\bf R} {\bf A}^{-1} {\bf f} = {\bf T} {\bf f}. \label{matrix_equation}\end{equation}  Here,  ${\bf T}= {\bf R} {\bf A}^{-1}$ is the so-called transfer matrix and ${\bf R} \in \mathbb{R}^{L \times N}$  denotes a restriction matrix picking the skin potential (voltage) values  at the electrode locations $e_1, e_2, \ldots, e_L$ on  $\partial \Omega$ and defining the zero potential level, e.g., as the sum of the entries of ${\bf y}$. If the $\ell$-th electrode is  placed at the $i_\ell$-th node then $R_{\ell,i_\ell} = 1 - 1/L$, $R_{\ell,i_j} = -1/L$, if $\ell \neq j$, and  $R_{\ell,j} = 0$, if the $j$-th node is not associated with an electrode. 
\label{transfer_matrix}

\subsection{Dipolar sources}

The present  forward approach enables computation of the potential field corresponding to any primary current distribution in $H(\hbox{div})$. In this paper,  we concentrate on the piecewise linear and quadratic bases of $H(\hbox{div})$ \cite{ainsworth2003} assuming that the potential field $u$ is spanned by a piecewise linear nodal basis and that the FE mesh is composed of tetrahedral elements. For the importance of the dipole source in EEG forward computations,  we associate each basis function  with a dipolar source. 

\begin{figure}[h]
\begin{center} 
\begin{minipage}{5cm}
\begin{center}
    \includegraphics[width=3.5cm]{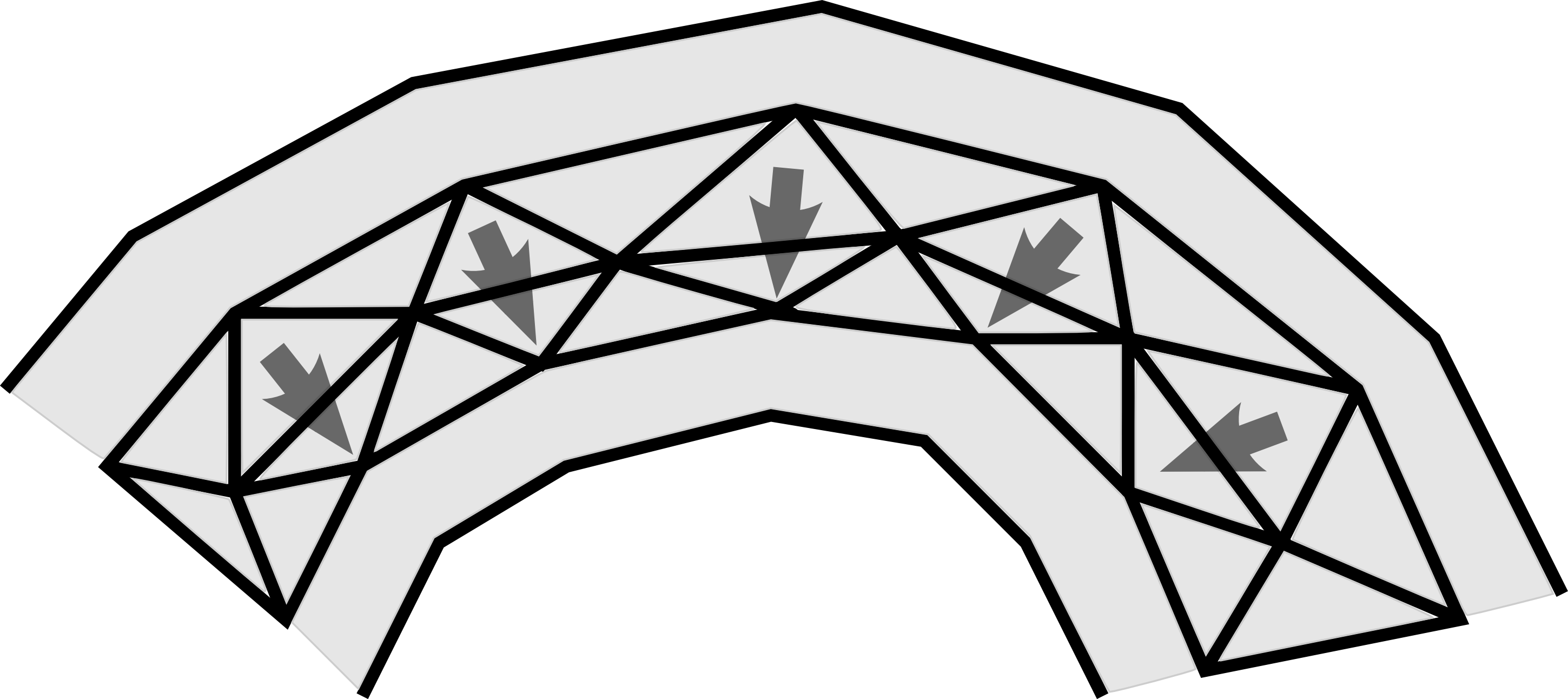} \\
Face intersecting (FI)
\end{center}
\end{minipage}  
\begin{minipage}{5cm}
\begin{center}
\includegraphics[width=3.5cm]{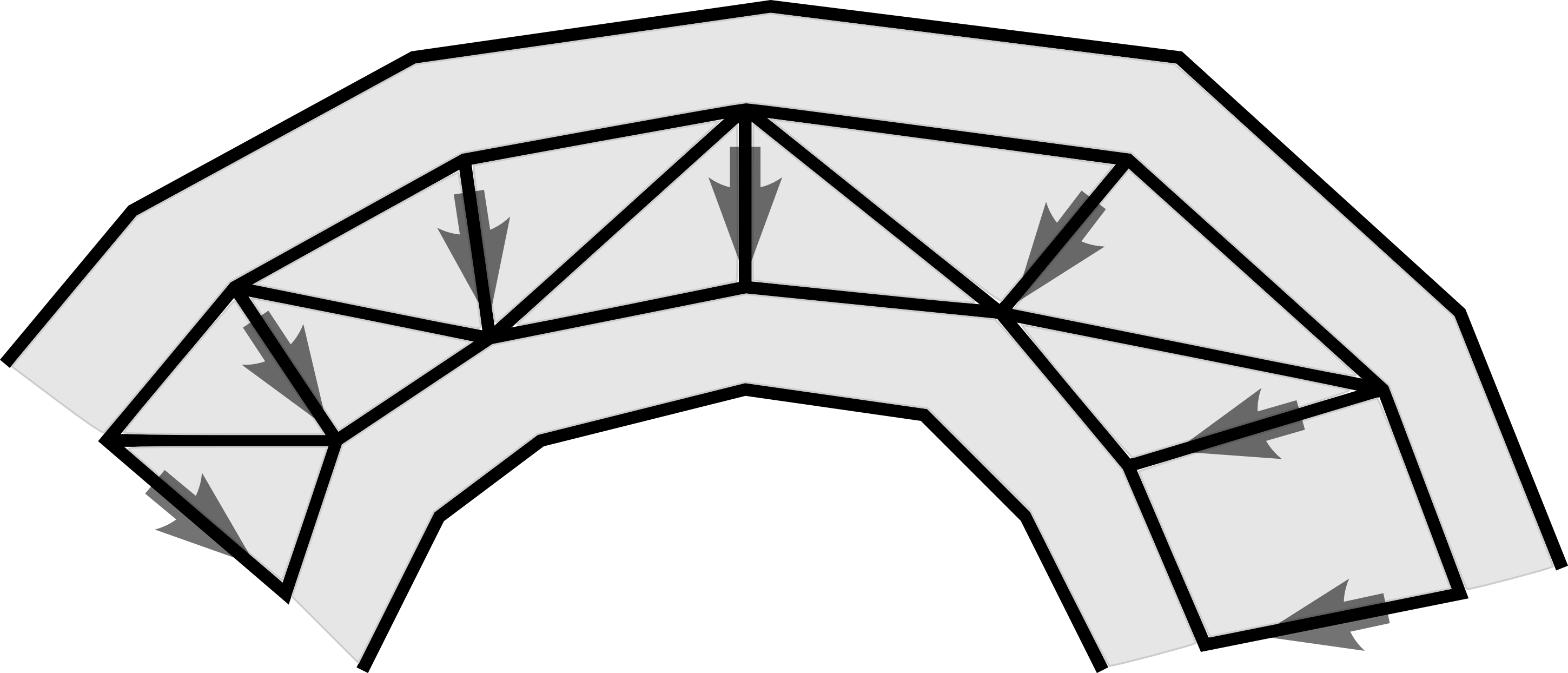} \\ 
Edgewise (EW)
\end{center}
\end{minipage} 
\end{center}
    \caption{A schematic picture showing  the face intersecting (FI)  and edgewise (EW) sources (linear and quadratic  $H$(div) basis functions, respectively (Appendix)) oriented roughly parallel to the inward normal of the grey matter layer (grey) to approximate the function of the pyramidal neurons  \cite{CHW:Creu62,CHW:Sch90}. This potential future mesh optimization approach was not applied in this study.   \label{cortex}}
\end{figure}

Defining the  dipolar moment of the basis function $\vec{w}$  as the integral  $\vec{q}_{\vec{w}} = \int_\Omega \vec{w} \; dV$, it follows that  $\vec{q}_{\vec{w}}$ is determined by the difference of two mesh nodes $P_i$ and $P_j$  as given by (see Appendix)  
\begin{equation} {\vec  q}_{\, \vec{w}} =    \frac{\vec{r}_{P_j}  - \vec{r}_{P_i}}{\| \vec{r}_{P_j}  - \vec{r}_{P_i} \|},
\end{equation}
where $\vec{r}_{P_i}$ and $\vec{r}_{P_j}$ denote the position vectors of $P_i$ and $P_j$.  
Moreover, the right-hand side matrix ${\bf G}$ of the discretized potential equation (\ref{jees}) is of the form (Appendix)
\begin{equation}
G_{\psi,\vec{w}}  = \int_\Omega  \psi (\nabla \cdot {\vec w} )  d V = \frac{s_{\{ \psi, P_j \}} \! - \! s_{\{ \psi, P_i\}} }{\| \vec{r}_{P_j}  - \vec{r}_{P_i} \|}
\end{equation}
with 
\begin{equation}
s_{\{ \psi,  P\}} = 
\left\{ \begin{array}{ll} 1, &  \hbox{ if } \psi \hbox{ corresponds to node } P, \\ 
0, & \hbox{ otherwise.} 
\end{array} \right. 
\end{equation}
That is, the load vector ${\bf f} = {\bf G} {\bf x}$ of (\ref{jees}) for a single  basis function ${\vec{w}}$ has a non-zero entry at the $i$-th and $j$-th node, and is zero otherwise. It follows that the natural choice for the position $\vec{r}_{\, \vec{w}}$ of a dipolar source is the midpoint of $P_i$ and $P_j$ \cite{bauer2015}
\begin{equation}
\vec{r}_{\, \vec{w}}   =  \frac{1}{2}(\vec{r}_{P_i}  + \vec{r}_{P_j}).
\label{pikkumyy2}
\end{equation} 
A straightforward calculation (Appendix) shows, that for a linear $H(\hbox{div})$ basis function, $P_i$ and $P_j$ are located on the opposite sides of the common face in an adjacent pair of tetrahedra. In the quadratic case, $P_i$ and $P_j$ are connected by an edge. Hence, if interpreted as dipolar sources, linear and quadratic  basis functions have face  intersecting (FI) and edgewise (EW) orientations, respectively. Motivated by this fact, we study these two orientation modes separately. Figure  \ref{cortex} shows an example of geometry-adaptation where the FI and EW sources have been organized roughly normal to the grey matter layer to  approximate the principal orientation of the pyramidal neurons \cite{CHW:Creu62,CHW:Sch90}. 

\subsection{Interpolation Methods}

This section describes the Position Based Optimization (PBO)  and Mean Position/Orientation (MPO) method, which interpolate a given dipole with position $\vec{r}$ and dipole moment $\vec{p}$ through a sum of dipolar FI and EW sources, i.e., $\vec{r} \approx \sum_{\ell = 1}^L c_\ell \vec{r}_{\, \vec{w}_{\ell}}$ and $\vec{p} \approx \sum_{\ell =1}^L c_\ell \vec{q}_{\, \vec{w}_{\ell}}$. In the PBO method, the preference is on $\vec{p}$ over $\vec{r}$, while in MPO, both are given an equal importance.  

\subsubsection{Position Based Optimization}
\label{pbo}

The PBO strategy \cite{bauer2015} finds the coefficient vector  ${\bf c} = (c_1,c_2, \ldots,c_L)$ as the  solution of 
\begin{equation}
\min_{{\bf c}} \, \sum_{\ell = 1}^L c_\ell^2 \omega_\ell^2 \quad \hbox{subject to} \quad \m{ Q}  \v{c} = \v{p},
\label{eqn:PositionBasedOptimization}
\end{equation}
where parameter $\omega_\ell = \| \vec{r}_{\vec{w}_{\ell}} - \vec{r} \|_2$ is a weighting coefficient and  $\m{Q} = (\v{q}_{\vec{w}_{1}},\v{q}_{\vec{w}_2}, \ldots, \v{q}_{\vec{w}_L})$. Constraint $\m{ Q}  \v{c} = \v{p}$ guarantees that the orientations of the interpolated and actual dipole will coincide.  For the convexity of $\sum_{\ell = 1}^L c_\ell^2 \omega_\ell^2$, the solution of  (\ref{eqn:PositionBasedOptimization}) can be obtained by applying the method of Langrangian multipliers, which yields a uniquely solvable linear system \begin{equation}  \left( \begin{array}{cc}  \v{D} & {\m{Q}}^T \\ \v{Q} & \v{0}  \end{array} \right) \left( \begin{array}{cc} \v{c} \\ \v{d} \end{array} \right)  = \left( \begin{array}{cc}  \v{0} \\ \v{p} \end{array} \right) \label{pbo}\end{equation} with a  diagonal matrix  $\v{D} = \hbox{diag}( \omega_1^2 , \omega_2^2, \ldots ,  \omega_L^2)$ and an auxiliary multiplier vector $\v{d}=(\lambda_1, \lambda_2, \lambda_3)$. The resulting total number of interpolation conditions $(L + 3)$ is the number of rows in the matrix of (\ref{pbo}). 

\subsubsection{Mean Position/Orientation Method}

The goal in the present MPO method is to choose ${\bf c} = (c_1, c_2, \ldots, c_L)$ so that the conditions 
{\setlength\arraycolsep{2 pt} \begin{eqnarray} 
\label{st_venant_2}
\vec{p} & = & \sum_{\ell=1}^L c_{\ell } \vec{q}_{\vec{w}_\ell}, \nonumber\\
\vec{0}  & = &  \frac{1}{\alpha} \sum_{\ell= 1}^L c_{\ell } \vec{q}_{\vec{w}_\ell}   [  (\vec{r}_{\vec{w}_\ell}   - \vec{r} ) \cdot \vec{e}_j]  \quad \hbox{for} \quad j = 1, 2, 3, \label{st_venant_1}
\end{eqnarray}}
are satisfied. The first one of these is the orientation constraint, and the second one requires that the average position of the dipolar moments is that of the given  dipole in each Cartesian direction $\vec{e}_j$  for $j = 1, 2, 3$.  The parameter $\alpha$ is a uniform mesh-based reference distance given a value which is at least double (here three times) the length of the  longest edge in the FE mesh. In order to minimize accumulation of numerical errors,  the least-squares solution of  (\ref{st_venant_1}) with a minimal $\ell^2$-norm is produced via 
\begin{equation}
{\bf c}  = {\bf M}^\dag {\bf b} \quad \hbox{with} \quad
{\bf M} = 
\left( \begin{array}{c}
{\bf Q }   \\
{\bf Q} {\bf P}_1  \\ 
    {\bf Q} {\bf P}_2  \\ 
  {\bf Q}  {\bf P}_3  
\end{array}  \right) 
\quad \hbox{and} \quad
{\bf b}  = \left( \begin{array}{c} {\bf p} \\
{\bf 0} \\
{\bf 0} \\
{\bf 0}
\end{array} \right), 
\end{equation}
where \begin{equation} 
{\bf P}_{j} =  \frac{1}{\alpha} \hbox{diag} \big( (\vec{r}_{\vec{w}_1} - \vec{r}) \cdot \vec{e}_j, (\vec{r}_{\vec{w}_2} - \vec{r}) \cdot \vec{e}_j, \ldots, (\vec{r}_{\vec{w}_L} - \vec{r}) \cdot \vec{e}_j \big)
\end{equation} for $j = 1, 2, 3$ and $ {\bf M}^\dag$ denotes the Moore-Penrose pseudoinverse of  ${\bf M}$. The total number of interpolation conditions  is 12. 

\subsection{Reference Methods}

Instead of dipolar sources, the classical partial integration (PI) and St.\ Venant (SV) dipole estimation methods combine monopolar electric potential distributions (single-node sources) to  approximate a given dipole. Of these, the PI is suitable for extremely focal dipole estimation.  Otherwise, an improved accuracy and robustness can be obtained with SV.

\subsubsection{Partial Integration}

The PI approach \cite{CHW:Yan91,weinstein2000}  approximates for the dipole moment $\vec{p}$ placed at point $\vec{r}$ inside $\Omega$  via the formula
{\setlength\arraycolsep{2 pt} \begin{eqnarray}
f_i & = & \! - \! \int_\Omega ( \nabla \cdot \vec{J}^p) \psi_i dV \! = \! \int_\Omega \vec{J}^p \cdot \nabla \psi_i dV  \! - \! \int_{\partial \Omega} \partial_{\bf n} \vec{J}^p \cdot \psi_i d S \nonumber\\ 
& = & \int_\Omega \vec{J}^p \cdot \nabla \psi_i dV =  \left\{ \begin{array}{ll} \vec{p} \cdot \nabla \psi_i|_{\vec{r}}, & \hbox{if $\vec{r}$ in support of $\psi_i$},  \\ 0, & \hbox{otherwise}, \label{pi_method} \end{array} \right.
\end{eqnarray}}
which follows from the assumption that the primary current density is of the form $\vec{J}^p = \vec{p} \, \delta_{\vec{r}}$ with $\delta_{\vec{r}}$ denoting a delta distribution \cite{CHW:Yan91}.  By utilizing (\ref{pi_method}) one can directly form the right-hand side vector ${\bf f}$ of (\ref{matrix_equation}). The resulting number of PI conditions is 4 which coincides with that of the non-zero vector entries in ${\bf f}$.


\subsubsection{St.\ Venant Method}

In the SV method \cite{CHW:Sch94,buchner1997,toupin1965,medani2015}, the dipole moment $\vec{p}$ at $\vec{r}$ is approximated  via  monopolar loads  $m_0,m_1,m_2,\dots,m_L$ placed at the FE mesh nodes $\vec{r}_0, \vec{r}_1,\vec{r}_2, \ldots, \vec{r}_L$ of which $\vec{r}_1,\vec{r}_2, \ldots, \vec{r}_L$ share an edge with the node $\vec{r}_0$ closest to $\vec{r}$. The net effect of the monopoles is arranged to approximately match that of the dipole via the conditions 
{\setlength\arraycolsep{2 pt} \begin{eqnarray}
0 & = & \sum_{i=0}^L m_i \nonumber\\
\frac{1}{\alpha}\vec{p} & = & \sum_{i = 0}^K \frac{m_i}{\alpha} (\vec{r}_i  - \vec{r} ) \nonumber\\
0 & = & \sum_{i = 0}^L \frac{m_i}{\alpha^2} [(\vec{r}_i  - \vec{r}) \cdot \vec{e}_j]^2 \quad \hbox{for} \quad  j=1,2,3, 
\label{st_venant}
\end{eqnarray}}
in which the reference distance $\alpha$ is at least double the length of the longest edge in the FE mesh (here three). From top to bottom, the conditions in (\ref{st_venant}) correspond to 
the conservation of the charge, the approximation of the dipole moment and the  suppression of higher order moments. 
The standard way to compute the load vector $\v{m} = (m_1,m_2,\ldots,m_L)$ is to determine the regularized least-squares estimate $\v{m} = ( {\bf P}^T {\bf P} + \lambda {\bf D})^{-1} {\bf P}^T {\bf b}$, where  \begin{equation} 
  {\bf P} =    
\left( \begin{array}{cc} {\bf P}_1 \\ {\bf P}_2 \\ {\bf P}_3 
\end{array} \right) \quad \hbox{with} \quad 
\v{P}_j = \left( \begin{array}{ccc} 1 &  \cdots & 1 \\ 
{\alpha}^{-1}(\vec{r}_1  - \vec{r}) \cdot \vec{e}_j & \cdots & {\alpha}^{-1} (\vec{r}_L  - \vec{r}) \cdot \vec{e}_j \\
\alpha^{-2}[(\vec{r}_1  - \vec{r}) \cdot \vec{e}_j]^2  &  \cdots & \alpha^{-2}[(\vec{r}_L  - \vec{r}) \cdot \vec{e}_j]^2 
 \end{array} \right),
\end{equation}
matrix ${\bf D} = \hbox{diag}(\| \vec{r}_1 -\vec{r} \|^2, \| \vec{r}_2 -\vec{r} \|^2,  \ldots, \| \vec{r}_L -\vec{r} \|^2)$ is a regularization matrix multiplied by the regularization parameter $\lambda > 0$ (here $\lambda = 10^{-6}$), and
\begin{equation}
 {\bf b} = \left( \begin{array}{c} 
{\bf b}_1 \\ {\bf b}_2 \\ {\bf b}_3 
\end{array} \right) \quad \hbox{with} \quad \v{b}_j = \left( \begin{array}{cc}  0 \\  \alpha^{-1} p_j \\ 0 \end{array} \right).
\end{equation} 
The total number of SV conditions is 7.

\subsection{Source configurations}

\begin{figure}[h] \begin{scriptsize}
\begin{center}   \begin{minipage}{3.9cm} \begin{center}  (A)   
\includegraphics[width=2.2cm]{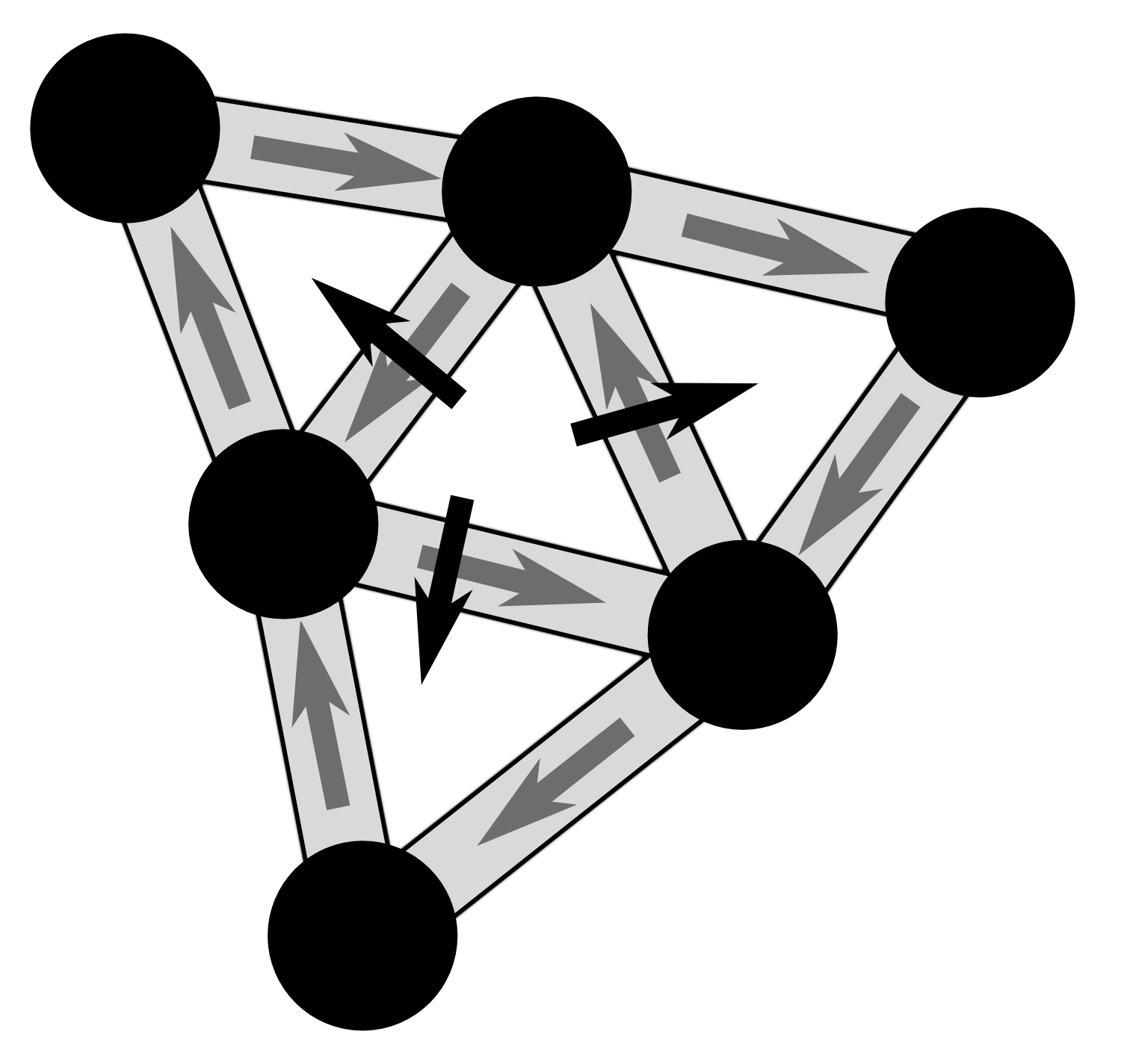}   \end{center} 
\begin{itemize} \item Elements that share a face with a given element \item 8 FE mesh nodes \item 22 dipolar sources \item 4 FI sources (black) \item 18 EW sources (grey) \end{itemize}  \end{minipage}
     \begin{minipage}{3.9cm} \begin{center} (B)  \includegraphics[width=2.2cm]{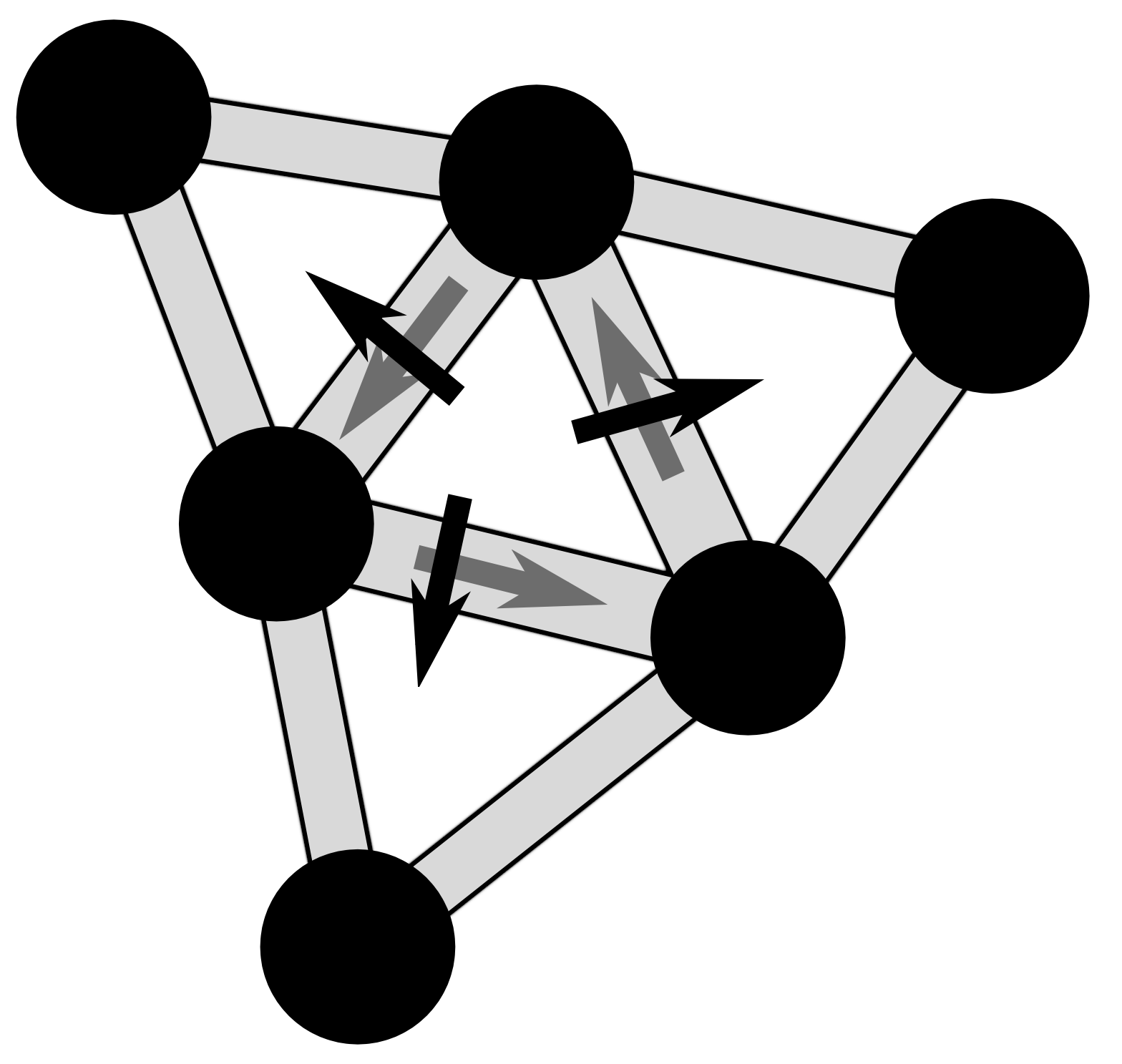} \end{center}  \begin{itemize} \item Elements that share a face with a given element \item 8 FE mesh nodes \item 10 dipolar  sources \item 4 FI sources (black) \item 6 EW sources (grey) \end{itemize} \end{minipage}
  \begin{minipage}{3.9cm} \begin{center}  (C) \includegraphics[width=2.2cm]{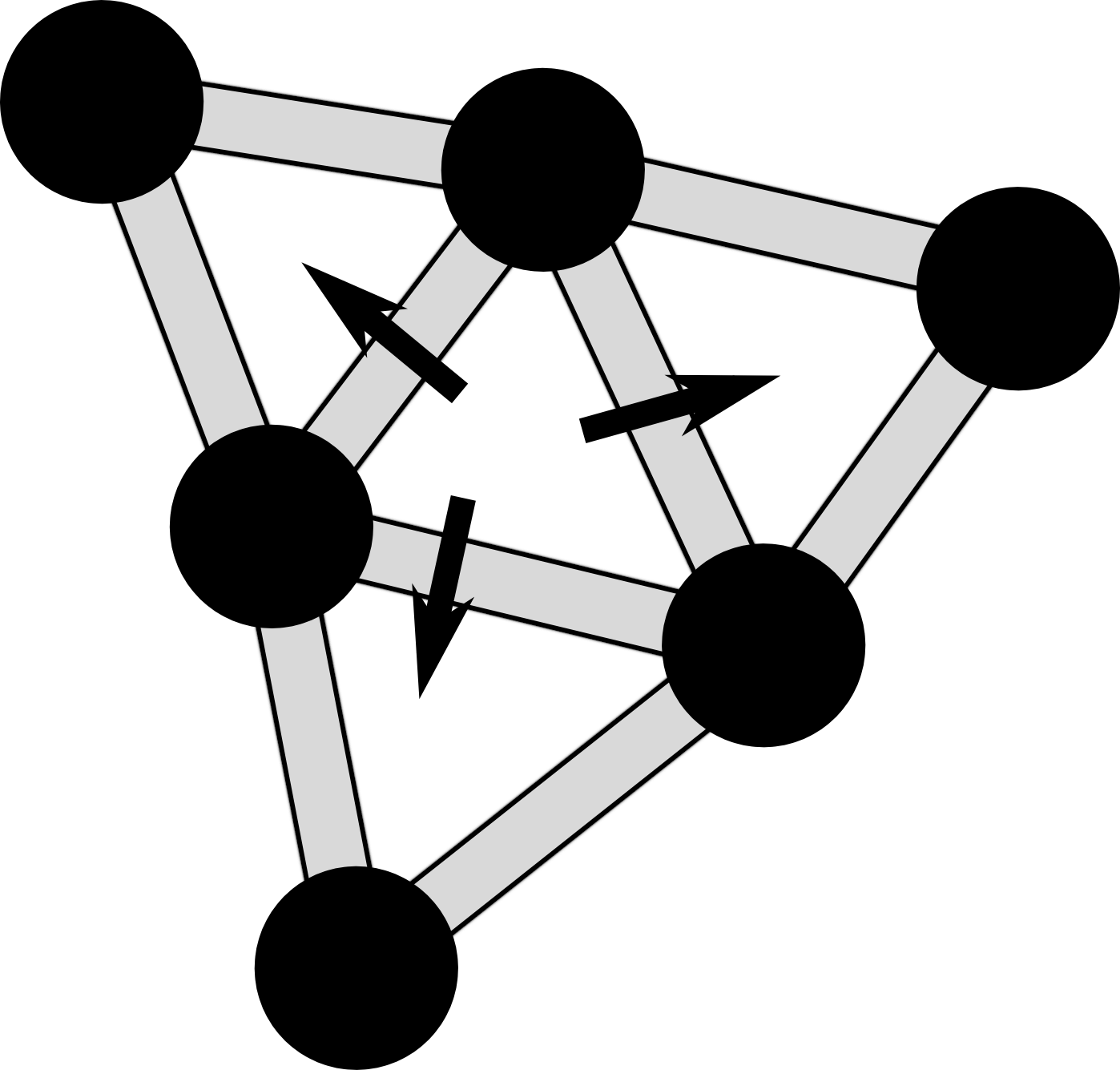} \end{center}  \begin{itemize} \item Elements that share a face with a given element \item 8 FE mesh nodes \item 4 dipolar  sources \item 4 FI sources (black) \item 0 EW sources (grey) \end{itemize}    \end{minipage}  \\ \vskip0.1cm
\begin{minipage}{3.9cm} \begin{center} (D) \includegraphics[width=1.5cm]{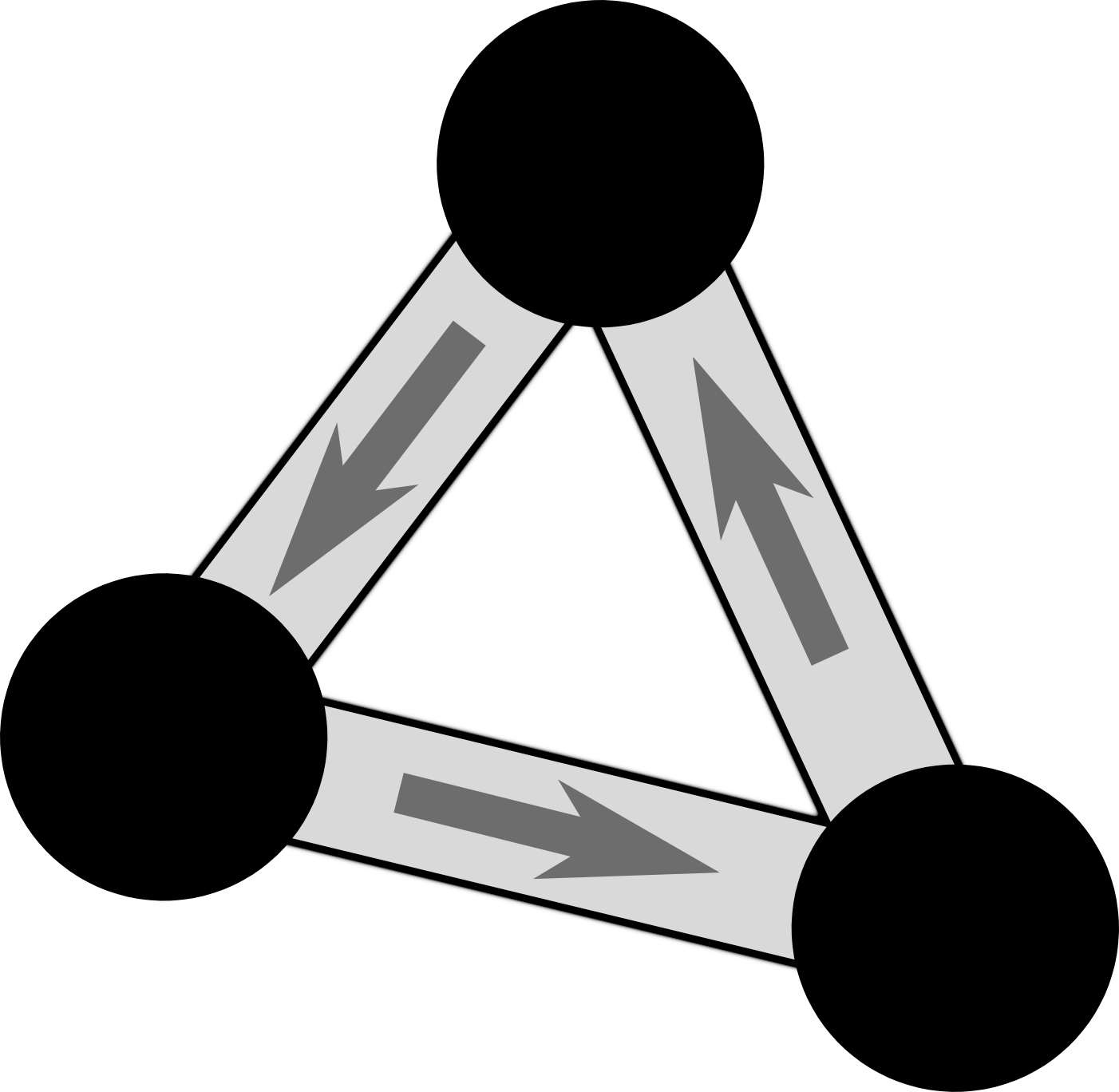} \\ \end{center}  \begin{itemize} \item A single element  \item 4 FE mesh nodes \item 6 dipolar  sources \item 0 FI sources (black) \item 6 EW sources (grey)  \end{itemize}   \end{minipage} 
 \begin{minipage}{3.9cm} \begin{center} (E)  \includegraphics[width=1.8cm]{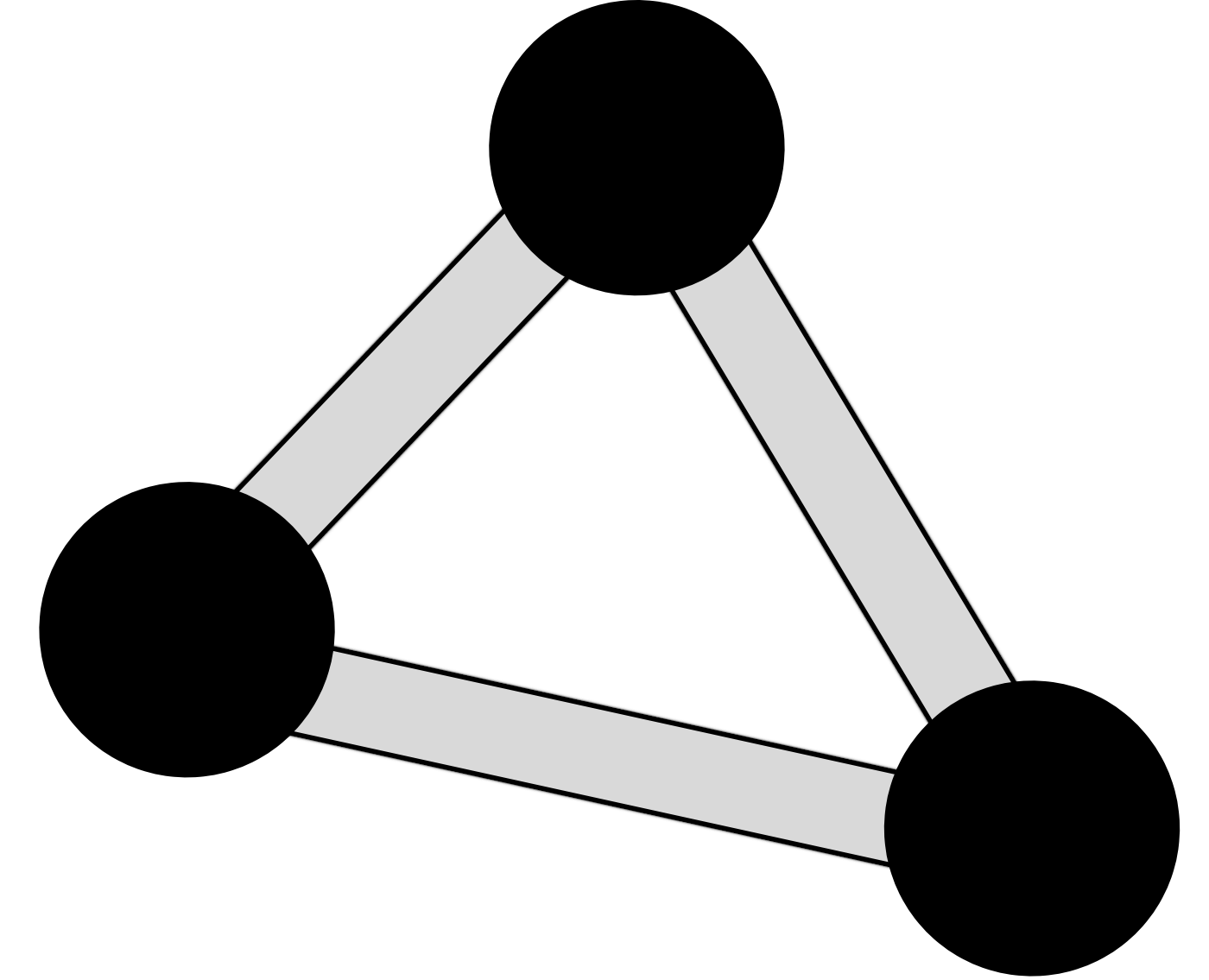} \end{center}  \begin{itemize} \item A single element \item 4 FE mesh nodes \item 4 monopolar  sources \\ \mbox{} \\ \mbox{} \\ \end{itemize} \end{minipage} 
 \begin{minipage}{3.9cm} \begin{center} (F) \includegraphics[width=2.15cm]{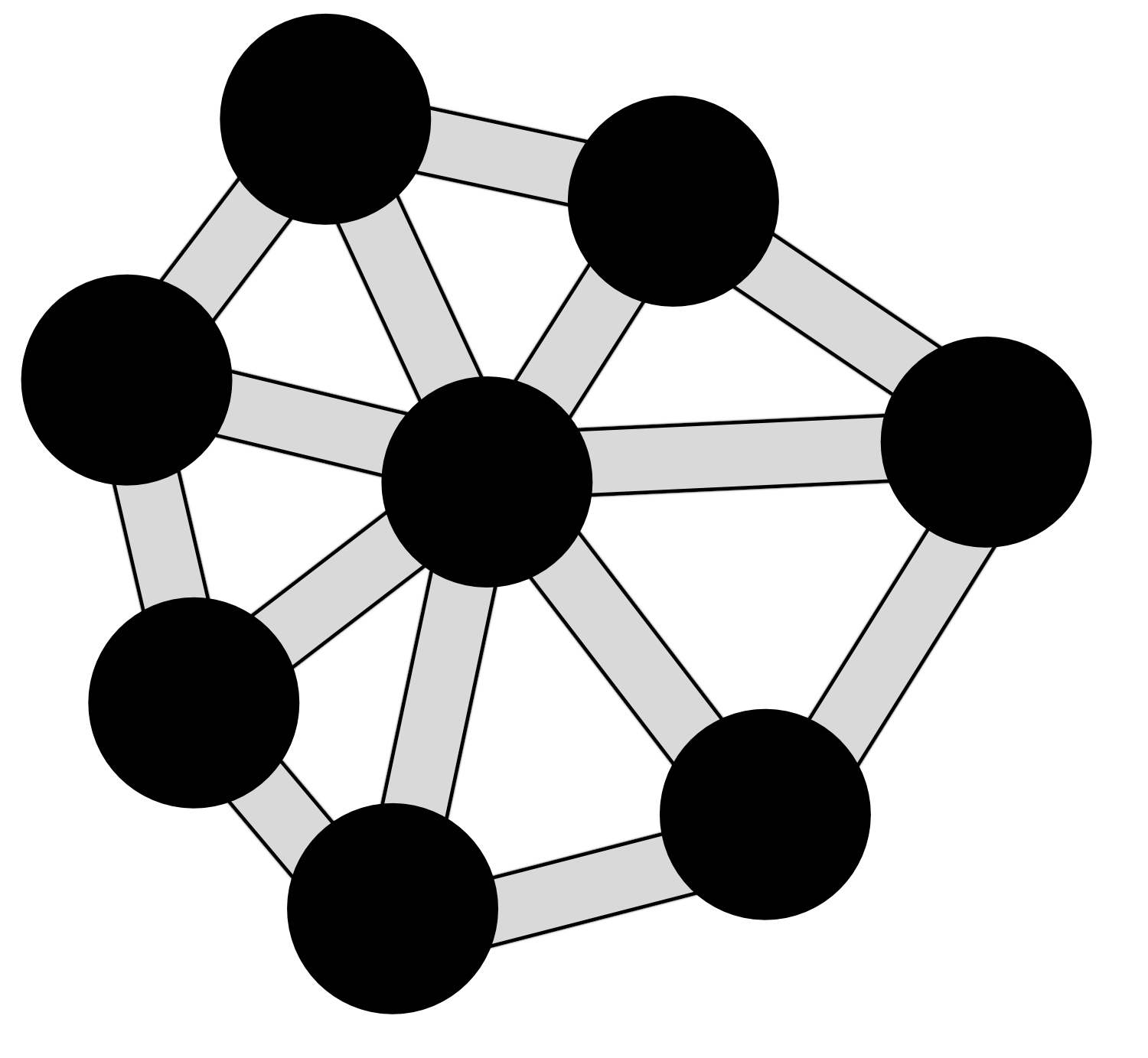} \end{center}  \begin{itemize} \item Elements sharing a given node \item  $\approx$16--27 FE mesh nodes \item   $\approx$16--27  monopolar  sources \\ \mbox{} \\ \mbox{} \\ \mbox{} \\ \end{itemize}   \end{minipage} 
\end{center}
\end{scriptsize}  
\caption{Detailed descriptions and schematic 2D visualizations of the 3D source  configurations (A)--(F). In the dipolar configurations (A), (B) and (C), the number of related mesh nodes (black dots) is $8$, and in (D), it is 4. The monopolar reference configurations (E) and (F) include 4 and $\approx$16--27 nodes, respectively. Configurations (A)--(D) can be applied in the PBO and MPO interpolation and (E) and (F) in PI and SV, respectively. With regard to focality (practical applicability),  (A)--(F) can be organized from the most to the least focal one as follows: 1.\ $\{D, E \}$, 2.\  $\{ A, B, C \}$ and 3.\ $\{ F \}$. It is important that all nodal  basis functions associated with a source configuration belong to the compartment of the grey matter in order to avoid modeling errors due to conductivity discontinuities. }
  \label{source_configuration}
\end{figure}

Figure \ref{source_configuration} describes four configurations (A)--(D) of dipolar sources utilized in the PBO and MPO methods and also two monopolar configurations (E) and (F)  corresponding to the PI and SV reference approaches, respectively. Since for each configuration, it is the number of nodes which determines the focality (practical  applicability),  (A)--(F) can be organized from the most to the least focal one as follows: 1.\ $\{D, E \}$, 2.\  $\{ A, B, C \}$ and 3.\ $\{ F \}$. In order to achieve an appropriate modeling accuracy, it is important that all nodal basis functions associated with a source configuration belong to the compartment of the grey matter.  Otherwise, errors due to conductivity  discontinuities may occur. 

\begin{table}[h]
  \caption{Radii and conductivies for the Stok model \cite{stok1987,CHW:Mun93,CHW:Kyb2005} (for adult head) formed by four concentric spherical compartments.}
  \label{conductivity}
  \begin{indented}
\item[]
\begin{tabular}{@{}lcccc}
\br
      Compartment & Scalp &  Skull &  CSF &  Brain    \\
      \mr
      Outer shell radius (mm) & 92 & 86 &  80 & 78 \\
      Conductivity (S/m) & 0.33 & 0.0042 & 1.79 & 0.33  \\
\br
  \end{tabular}
\end{indented}
\end{table}

\begin{figure}[h]
\begin{center} 
\begin{minipage}{5cm}
\begin{center}
    \includegraphics[width=3.2cm]{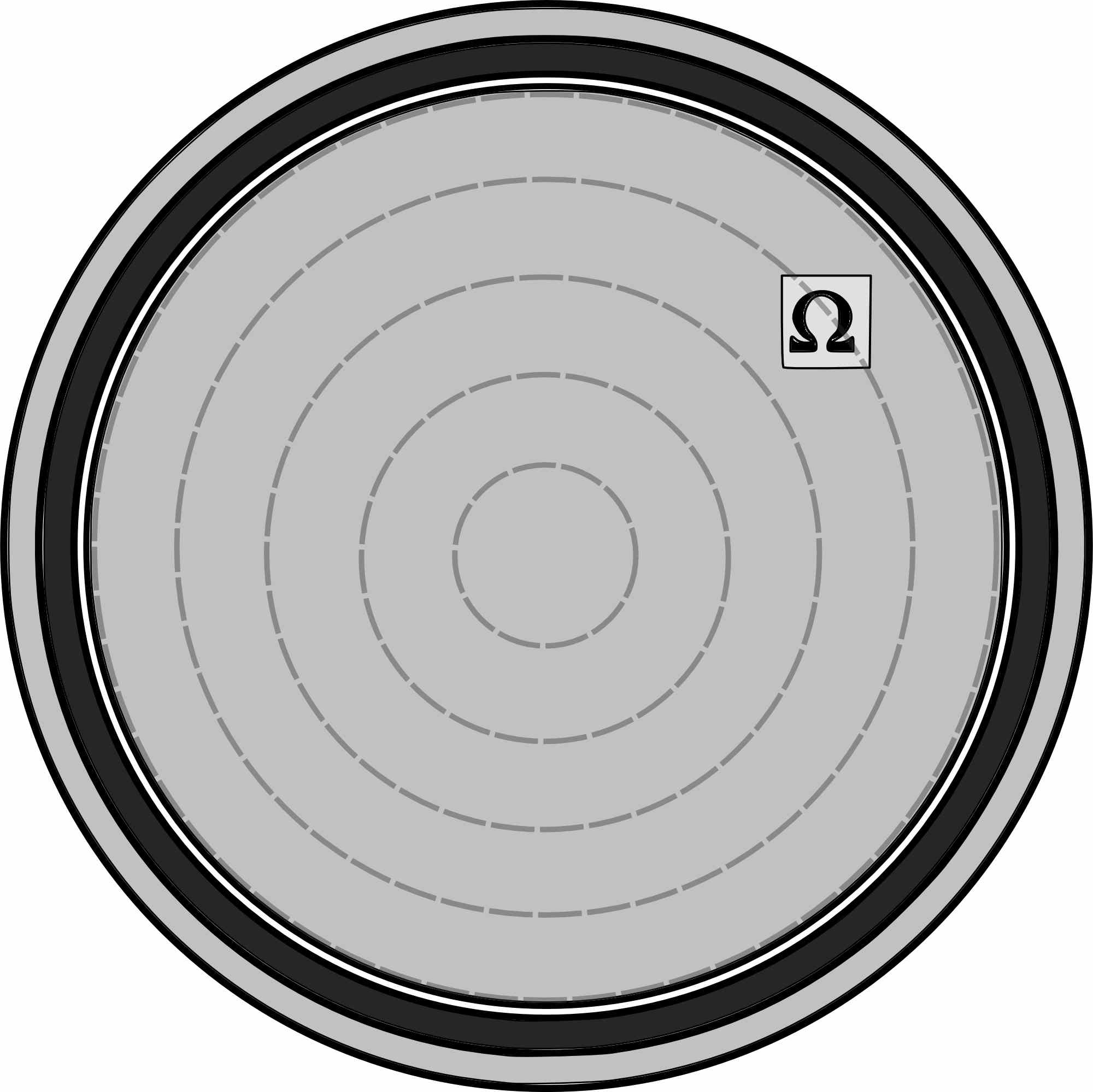} \\ Stok model
\end{center}
\end{minipage}  
\begin{minipage}{5cm}
\begin{center}
\includegraphics[width=3.2cm]{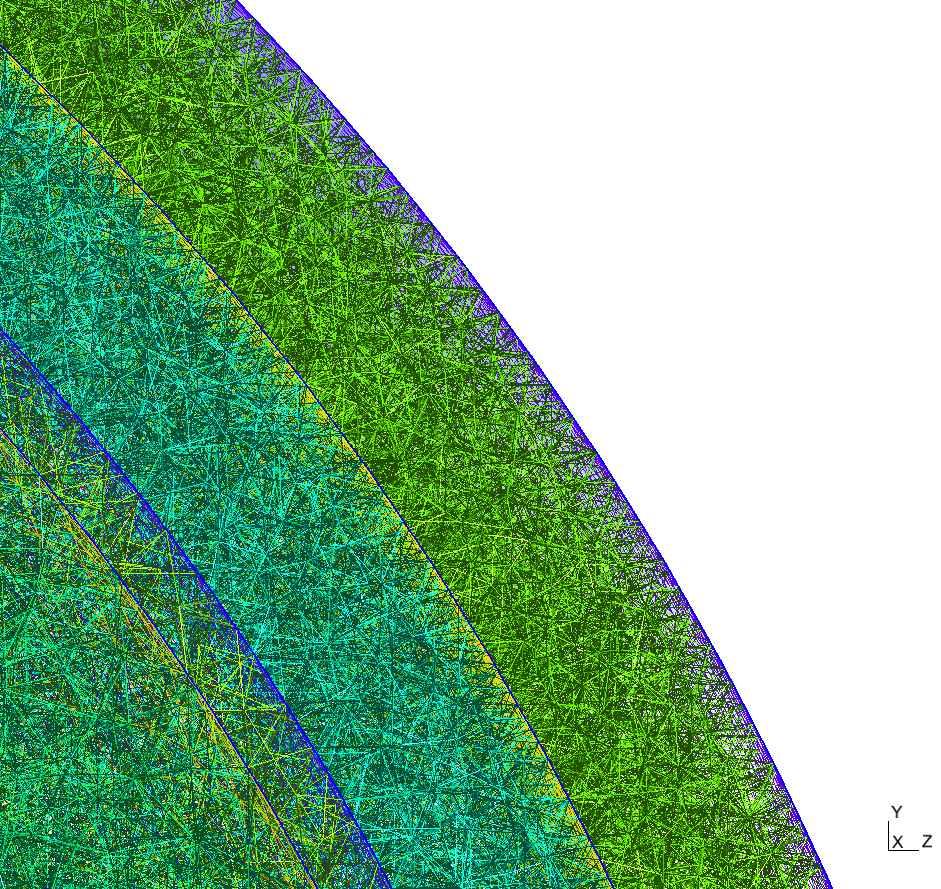} \\ Finite element mesh
\end{center}
\end{minipage} 
\end{center}
    \caption{Left: A schematic illustration of the Stok model, the domain $\Omega$ of this study. The compartments are from the innermost to the outernmost one correspond to the brain, cerebrospinal fluid (CSF),  skull, and skin. The dipole locations distributed on spherical surfaces  (20, 40, 60, 80, and 99\% eccentricity) are illustrated by the dashed light grey circles within the brain compartment. Right: A cross-section of the 3D finite element mesh based on the Stok model.  \label{figure_stok_model}}
\end{figure}

\begin{table}[h]
  \caption{The combinations of interpolation method (PBO/MPO/PI/SV)  and source configuration ((A)--(F)) explored in the numerical experiments. The number of linear interpolation conditions is also included for each method.}
  \label{combinations}
  \begin{indented}
\item[]
\begin{tabular}{@{}lccccccc}
\br
   Method   &  Conditions & (A) & (B) &  (C)  &  (D) & (E) & (F)    \\
      \mr
PBO &  7 & * & * & * & *  &  & \\
 MPO  & 12 & * &  * & * & *   & &  \\
  PI  &  4 &    &   &    &   & *&   \\ 
  SV  & 7 &    &   &    &   & & * \\ 
\br
  \end{tabular}
\end{indented}
\end{table} 

\subsection{Numerical Experiments} 

The numerical experiments were performed utilizing the  Stok model \cite{stok1987} (Figure \ref{figure_stok_model}) which substitutes spherical concentric compartments for the brain, cerebrospinal fluid (CSF), skull, and skin with radii 78, 80, 86, and 92 mm and conductivities of 0.33, 0.0042, 1.79, and 0.33 S/m, respectively (Table \ref{conductivity}). The electrode voltage vector ${\bf y}$ consisted of 200 entries corresponding to an even spherical spread of boundary points. The domain was discretized with a FE mesh consisting of 801,633 nodes and 4,985,234 tetrahedra. The mean size of a tetrahedron was 0.7 mm. The reason for choosing a spherical model was the existence of the analytical solution that enables validating the accuracy of the FEM approximation. 

The results yielded by the FEM were analyzed using the following relative difference and magnitude measures (RDM and MAG) expressed in percents:
{\setlength\arraycolsep{2 pt} \begin{eqnarray} \label{jepjep_1}
 \hbox{RDM}( \v{y}_{\hbox{\scriptsize ana}}, \v{y}_{\hbox{\scriptsize FEM}} )& = & \frac{100}{2} \left\|      \frac{ \v{y}_{\hbox{\scriptsize ana}}  } {\| \v{y}_{\hbox{\scriptsize ana}} \|_2 } -   \frac{ \v{y}_{\hbox{\scriptsize FEM}}  }     {\| \v{y}_{\hbox{\scriptsize FEM}} \|_2 }  \right\|_2, \\ \hbox{MAG}( \v{y}_{\hbox{\scriptsize ana}}, \v{y}_{\hbox{\scriptsize FEM}} ) & = &  100 \frac{\| \v{y}_{\hbox{\scriptsize FEM}} \|_2 }     { \| \v{y}_{\hbox{\scriptsize ana}} \| _2} - 100, \label{jepjep_2}
\end{eqnarray}}  
where $\v{y}_{\hbox{\scriptsize FEM}}$ and $\v{y}_{\hbox{\scriptsize ana}}$ correspond to FEM based and analytical forward simulations. Of these, the RDM estimates positional and directional  differences between analytical and numerical dipole approximation, and MAG measures magnitude differences. In the context of inverse source detection, RDM relates to location and orientation error and MAG the magnitude of the source.

First, the dipolar moments and positions of analytic dipoles were fixed to those of  FI and EW sources. The eccentricity (i.e.\ the relative distance from the center within the brain) values of 20, 40, 60, 80 and 99\% were covered.  A sample of 200 sources was created for each eccentricity value and the corresponding RDM and MAG distributions were investigated via box-plots \cite{mcgill1978}. In order to determine statistically significant mutual differences between the samples, we used the non-parametric Mann-Whitney U-test for the median with 95\% confidence interval \cite{mann1947}. In this context, a difference that can be systematic rather than totally random is considered significant. It is of note that the U-test is used here to provide complementary information of the statistical distributions and that significance does not refer to a practically or clinically meaningful differences. 

Subsequently, different interpolation schemes (Table \ref{combinations}) were explored  and also evaluated against the reference methods. In this experiment, dipoles with randomized source positions and orientations were to be simulated at the eccentricity of 99\%, which corresponds to a distance less than 0.8 mm to the outer brain surface. Since   numerical errors are known to increase along with the eccentricity \cite{CHW:Wol2007e}, this test is valid for a cortical thickness greater than 2 $\times$ 0.8 mm $=$ 1.6 mm, when each nodal basis function utilized in the interpolation is inside the grey matter compartment. Cortical thicknesses around 1.6 mm are commonly found in infants \cite{li2014}. 

\section{Results}
\label{results}

\begin{figure}[h]
\begin{footnotesize}
\begin{center} 
\begin{minipage}{10cm}
\begin{center}
   \includegraphics[width=10cm]{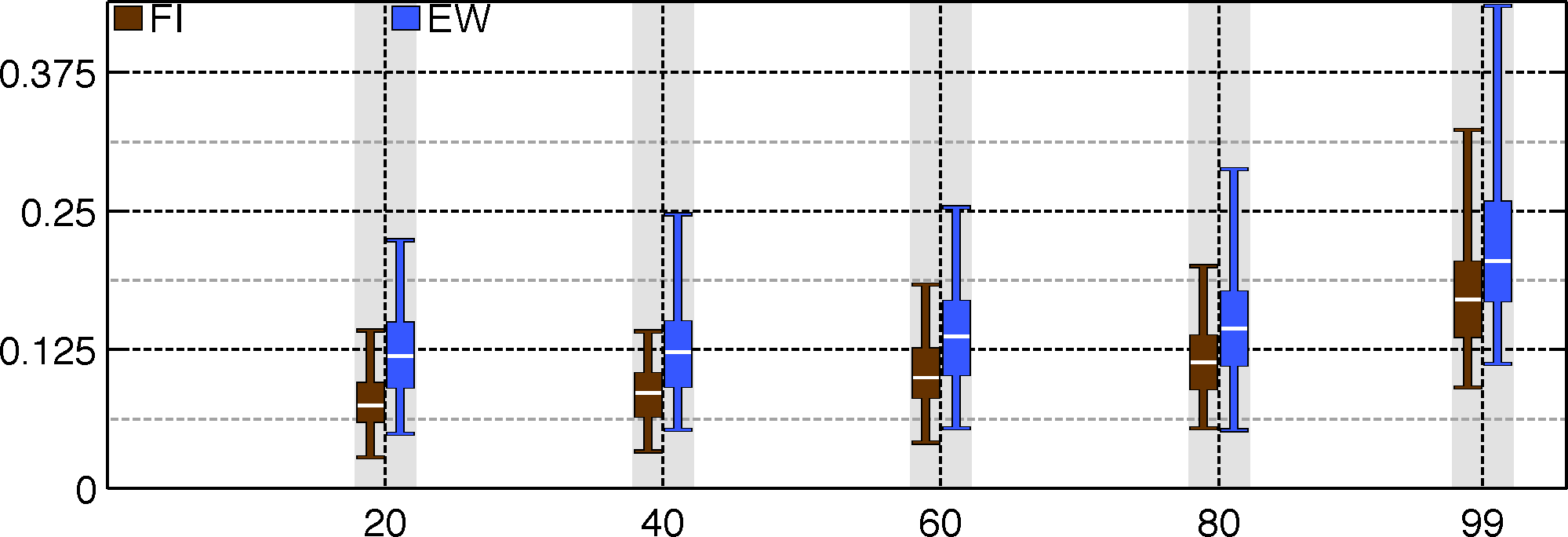}  \\ RDM 
\end{center}
\end{minipage}  \vskip0.1cm  \mbox{} \\
\begin{minipage}{10cm}
\begin{center}
   \includegraphics[width=10cm]{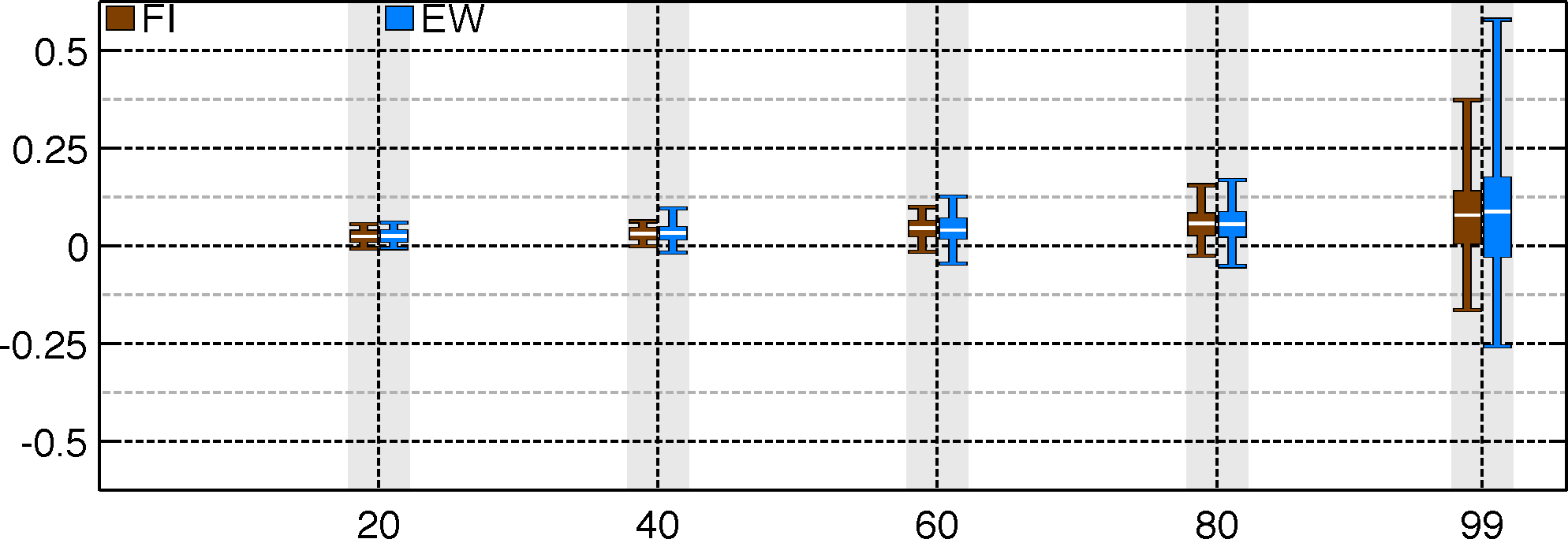}  \\ MAG 
\end{center}
\end{minipage}
\end{center}
\end{footnotesize}
  \caption{Relative difference measure (RDM in \%) and magnitude measure (MAG in \%)  for face intersecting (FI) and edgewise (EW) sources at eccentricities of 20, 40, 60, 80, and 99\%. Here, the dipolar moments and positions of analytic dipoles correspond to those of FI and EW sources. The box-plot bars show the median, the interval between the maximum and minimum, i.e., the total range (TR), and the interval between upper and lower quartile, known as the interquartile range (IQR) or spread. }
  \label{(i)}
\end{figure}

\begin{figure}[h]
\begin{footnotesize}
\begin{center} 
\begin{minipage}{10cm}
\begin{center}
   \includegraphics[width=10cm]{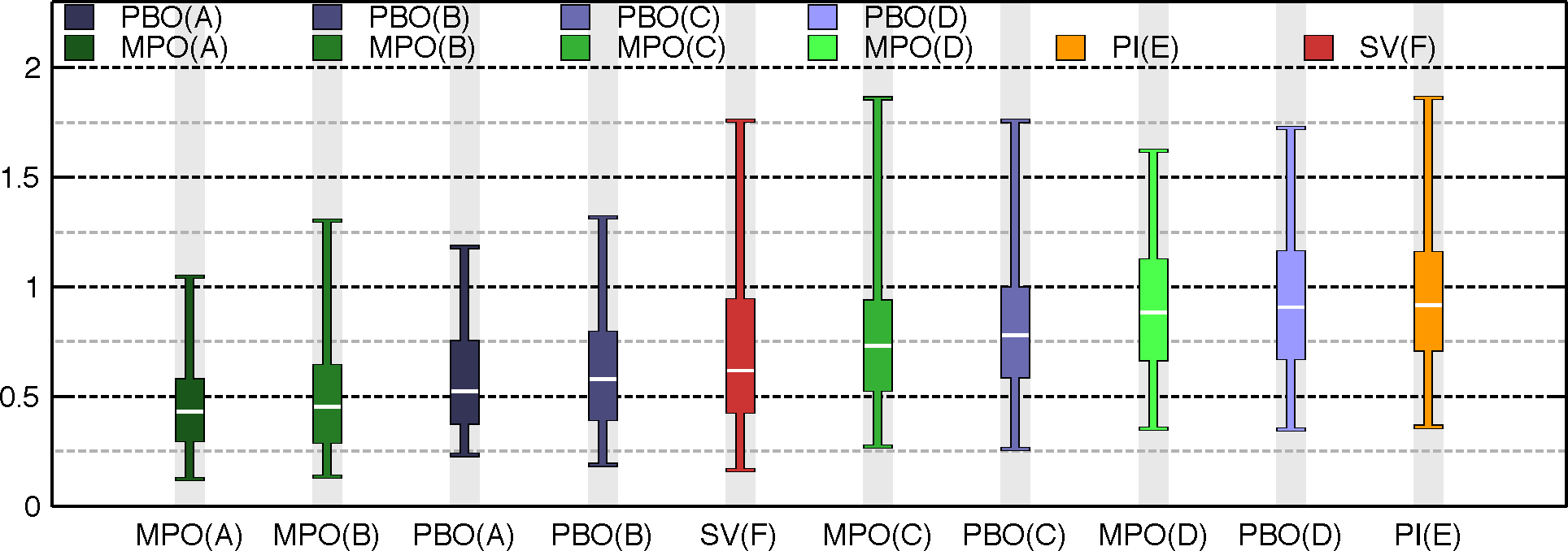}  \\ RDM  \vskip0.1cm \mbox{}  \\ 
\end{center}
\end{minipage} 
\begin{minipage}{10cm}
\begin{center}
   \includegraphics[width=10cm]{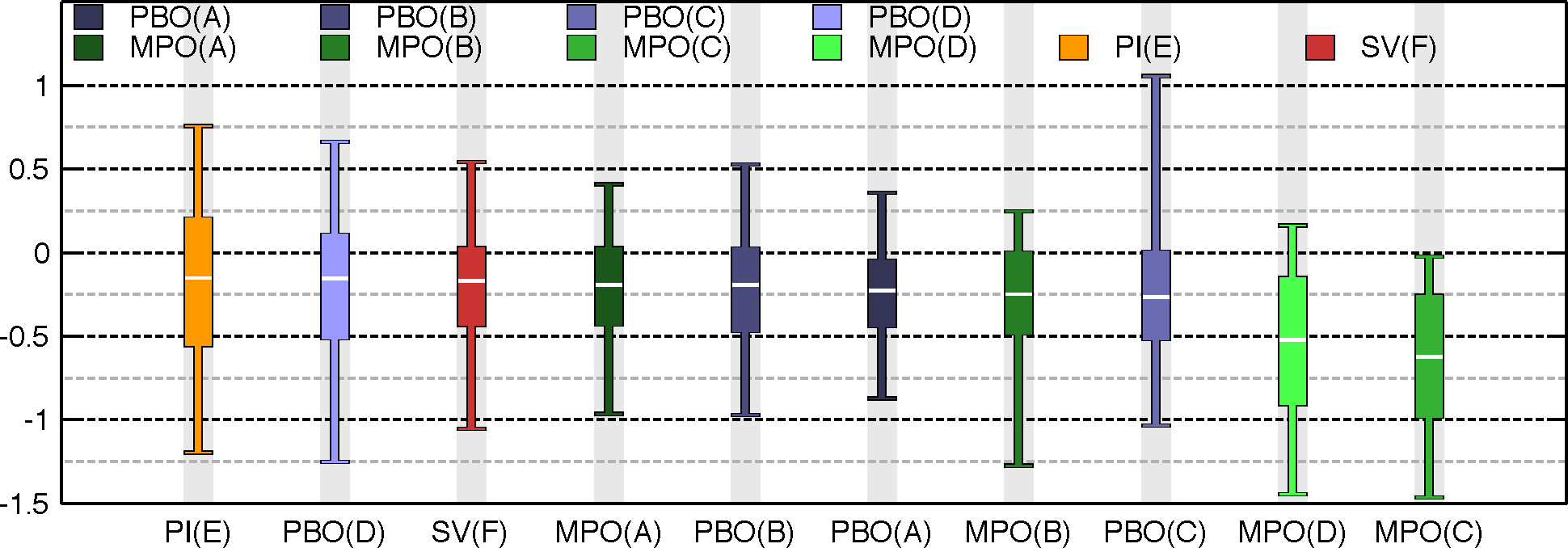} \\  MAG \\  \mbox{}
\end{center}
\end{minipage}  

\end{center}
\end{footnotesize}
  \caption{Relative difference measure (RDM in \%) and magnitude measure (MAG in \%) in interpolated approximation of randomized dipoles at the eccentricity of 99\%. The interpolation schemes have been ranked in an ascending order  from left (best) to right (worst) based on the absolute value of the median. The box-plot bars show the median, the interval between the maximum and minimum, i.e., the total range (TR), and the interval between upper and lower quartile, known as the interquartile range (IQR) or spread.}
  \label{(ii)}
\end{figure}

\begin{table}[t]
\caption{The results of the Mann-Whitney U-test with confidence interval 95\%. Significant mutual statistical differences in the median are marked with asterisk.}
\label{significance}
\begin{indented}
\item[]
\begin{tabular}{@{}lllllllllllll}
\br
& & & \centre{4}{PBO} & \centre{4}{MPO}  &  PI & SV   \\
& & & \crule{4}& \crule{4}  &  &   \\
& & & (A) & (B) & (C) & (D) &  (A) & (B) & (C) & (D) &   (E) & (F) \\
\mr
RDM & PBO & (A)  &         &      &   *  &   *  &   *  &  *  &   *  &  *  &  * &   * \\
& & (B)      &         &      &   *  &   *  &   *  &  *  &   *  &  *  &  * &    \\
& & (C)      &      *  &   *  &      &   *  &   *  &  *  &      &  *  &  * &   *\\
& & (D)      &      *  &   *  &   *  &      &   *  &  *  &   *  &     &    &   * \\
& MPO &(A)   &      *  &   *  &   *  &   *  &      &     &   *  &  *  &  * &   *\\
& &(B)       &      *  &   *  &   *  &   *  &      &     &   *  &  *  &  * &   *\\
& &(C)       &      *  &   *  &      &   *  &   *  &  *  &      &  *  &  * &   * \\
& &(D)       &      *  &   *  &   *  &      &   *  &  *  &   *  &     &    &   * \\
& PI &(E)    &      *  &   *  &   *  &      &   *  &  *  &   *  &     &    &   * \\
& SV & (F)   &      *  &      &   *  &   *  &   *  &  *  &   *  &  *  &  * &    \\
\mr
MAG & PBO & (A) &          &     &   &     &    &    &  * &  * &   &       \\
 &    & (B) &          &     &   &     &    &    &  * &  * &   &       \\
&    & (C) &          &     &   &     &    &    &  * &  * &   &       \\
 &   & (D) &          &     &   &     &    &  * &  * &  * &   &      \\
& MPO &(A)  &     &     &   &     &    &    &  * &  * &   &       \\
 &   &(B)  &          &     &   &  *  &    &    &  * &  * & * &      \\
 &   &(C)  &       *  &  *  & * &  *  & *  &  * &    &    & * &  *   \\
  &  &(D)  &       *  &  *  & * &  *  & *  &  * &    &    & * &  *   \\
& PI &(E)   &          &     &   &     &    &  * &  * &  * &   &      \\
& SV & (F)  &          &     &   &     &    &    &  * &  * &   &      \\
 \br
 \end{tabular}
 \end{indented}
\end{table} 

\begin{table}[t]
  \caption{A summary of the MPO and PBO interpolation schemes, based on the outcome of the numerical experiments. Based on this study, the configurations (B)--(D) are incompatible with  the  MPO method. For PBO, our preference ordering of the source configurations w.r.t.\ RDM and MAG has been given in the center column.}
  \label{summary}
  \begin{indented}
\item[]
\begin{tabular}{@{}llll}
\br 
& \centre{2}{Approach} &  \\ & \crule{2}  &\\
Feature & MPO              & PBO             & Reasoning     \\ 
\mr
Characteristics & Accuracy & Stability, Adaptability & \\
Accuracy preference & (A)                &  1.\ (A), 2.\ (B), 3.\ (C) & RDM results      \\
Focality preference & (A)  & 1.\ (D), 2.\ (A), (B), (C) & Number of nodes \\
Incompatible & (B), (C), (D) & - & MAG results\\
\br                    
  \end{tabular}
  \end{indented}
\end{table} 

The results of the numerical experiments have been included  in Figures \ref{(i)}--\ref{(ii)} and Table \ref{significance}. A summary with evaluations and recommendations based on the outcome of the  results  has been given in Table \ref{summary}. The detailed discussion  can be found in Section \ref{discussion}.  In all experiments, RDM and MAG  were below 2.0 and 1.5\%, respectively, indicating that an overall appropriate  level of accuracy had been achieved. The lowest RDM and MAG maxima, below 0.4  and 0.6\%, respectively,  were obtained with  FI and EW sources at their positions and orientations (Figure \ref{(i)}). The FI mode was found to be overall superior to the EW version; based on the U-test, the mutual differences in median were significant with respect to RDM  but not with respect to MAG. 

\subsection{RDM for interpolation schemes}

For both MPO and PBO interpolation approach, the smallest RDM median at the 99\% eccentricity was achieved by the source configuration (A) with a significant difference to the reference methods PI (E) and SV (F) (Figure \ref{(ii)} and Table \ref{significance}). The RDM median of MPO (A) was found to be significantly smaller than that of PBO (A). Furthermore, the median-based ranking of the RDM (Figure \ref{(ii)}) together with the outcome of the U-test  (Table \ref{significance}) suggest that, for both MPO and PBO, the  difference between configurations (A) and (B) is  negligible, (A) and (B) are superior to both (C) and (D), and (C) is preferable over (D). 

\subsection{MAG for interpolation schemes}

MPO (C) and MPO (D)  were observed  to be significantly inferior to all other investigated methods with respect to the median of MAG (Figure \ref{(ii)} and Table \ref{significance}). Additionally, based on the U-test, the MAG  median of MPO (B) differed  significantly from that of PI (E) and PBO (D), which were the two first methods in the median-based ranking (Figure \ref{(ii)}). Otherwise, the MAG median differences were found to be insignificant. 

\section{Discussion}
\label{discussion}

In this study,  divergence conforming $H$(div) basis functions \cite{monk2003,ainsworth2003,solin2003} were utilized in a finite element based bioelectromagnetic source modeling, focusing on its application in EEG  \cite{niedermeyer2004,brazier1961,braess2001,hamalainen1993,CHW:Mun2012}. The primary source current field $\vec{J}^P$  (Section \ref{forward_model}) was discretized as a superposition of piecewise linear and quadratic basis functions. For a tetrahedral finite element (FE) mesh, those can be associated with face intersecting \cite{pursiainen2011,pursiainen2012b,bauer2015} and edgewise dipolar currents, respectively. The resulting  divergence $\nabla \cdot \vec{J}_h^P$, i.e., the source term of the potential equation, is a piecewise linear function similar to the discretized potential field $u_h$. An arbitrary mathematical point-dipole was approximated via interpolation utilizing a set of FI and EW sources associated with at most eight nodes of the FE mesh. Position based optimization (PBO) \cite{bauer2015} and mean position/orientation  (MPO)  techniques together with source configurations (A)--(D) were compared  against each other and also with two widely used reference methods, the partial integration (PI) \cite{CHW:Yan91,weinstein2000} and St.\ Venant (SV)  \cite{CHW:Sch94,buchner1997,toupin1965,medani2015} approach. The results were analyzed statistically \cite{mcgill1978} via the relative difference and magnitude measures (RDM and MAG). 

With RDM and MAG below 2.0 and 1.5\%, respectively, all estimates of this study can be regarded as sufficiently accurate to be exploited in biomedical applications \cite{niedermeyer2004,hamalainen1993,CHW:Mun2012} in which errors due to other experimental factors (e.g.\ trigger accuracy) or methodological aspects  (geometry, interindividual conductivity, choice of inverse approach etc.) are present to at least a similar magnitude. The differences observed in the comparisons can, however, be relevant in a realistic modeling context, where the FE mesh is less homogeneous leading to less predictable error distributions.  Namely, within a realistic head geometry,  conductivity jumps can significantly reduce the estimation accuracy if, unlike in this study,  the source configuration  includes nodal basis functions supported in compartments other than the grey matter. That is,  if the simulated source is not restricted to the grey matter similar to the actual one, it will result in an erroneous   volume current $-\sigma \nabla u$ and thereby inexact electric potential $u$ (Figure \ref{contours}).  It is therefore obvious that the focality of the source configuration is extremely important in order to maximize the fit with grey matter. 

In agreement with previous reports \cite{bauer2015}, the most focal (two nodes) and accurate solution was provided by the non-interpolated  dipolar sources. The RDM and MAG obtained with optimized positions and orientations were substantially better, i.e. below  0.4  and 0.6\%, respectively, than when estimating an arbitrary dipole via interpolation. Furthermore, the FI orientation was found to be superior to the EW mode, indicating that the estimation accuracy obtained with the linear nodal basis decreases when the polynomial order of the source vector field is increased, that is  a natural consequence of the limited numerical resolution.  In practice, both the  interpolated and  non-interpolated approaches can be suitable for forward simulation purposes. The preferable choice between those two alternatives depends primarily on the associated methodology, such as inversion techniques, which can demand that the sources have to be placed in specific positions and/or orientations. If there are no limitations set by the context, then the non-interpolated (mesh-based) vector field of FI and EW sources can be considered as advantageous. To optimize such a field, it is desirable to design the FE mesh according to the {\em a priori} knowledge of the tissue structure, e.g., to approximate the function of pyramidal neurons evoking currents in the inward-pointing normal direction within the grey matter \cite{CHW:Creu62,CHW:Sch90}. At present, this represents a challenging task that will need to be studied in the future.  A further interesting topic is also to study finitely  supported currents  as a superposition of FI and EW sources, i.e., as piecewise quadratic vector fields. 

This study revealed that a combination of linear and quadratic polynomial fields was  advantageous over either of those separately;  the hybrid FI/EW eight-node configuration  (A) achieved the best overall estimation of accuracy and robustness.  Hence, (A)  may be optimal, if it fits appropriately to the grey matter compartment, i.e., if there are no conductivity jumps in the support of the corresponding set of nodal basis functions. Otherwise, one will need to adopt another solution, e.g., the single-element EW configuration (D), in order to avoid the forward errors caused by those jumps. 
Other important findings were that (A) and (B) achieved  better results than the previously studied FI source configuration (C), in particular PBO (C) \cite{bauer2015}, and that furthermore, the accuracy of the reference methods PI (E) and SV (F) could be attained or even surpassed, in some respects,  advocating the current eight-node approach as a preferable solution.  An important point  is that the present dipolar source configurations (A)--(D) are considerably more focal than (F) of approximately 16--27 nodes.

Since the number of dipolar sources in (A), (B) and (D) exceeds that of nodal (potential field) basis functions, utilizing a higher polynomial order of the potential field, and thus the number of associated degrees of freedom,  would be one way to improve the forward simulation accuracy. To prevent excessive growth of the resulting linear system, higher-order polynomial basis functions can be placed to a specifically selected subdomain or region of interest, e.g., to the grey matter compartment or only to its narrow parts. There are different ways to achieve this goal, for example, the scalar-valued fourth-order bubble function that is supported within a single tetrahedron \cite{ainsworth2003,solin2003}. 

The crucial technical difference between the PBO and MPO interpolation methods is that the orientation constraint is exact in the first approach, whereas in the second technique,  it holds only in the least-squares sense. The major reason for this latter limitation is that the more detailed positioning conditions of MPO do not allow adoption of the convex optimization approach utilized in PBO. MPO was found to be the superior approach for configuration (A). PBO worked well also for (B), (C) and (D), whereas MPO yielded  significantly larger magnitude errors for those configurations. Consequently, PBO seems to be preferable over MPO, when the associated linear system is overdetermined, i.e., when the number of interpolation conditions is greater than the source count. In practice, this means,  that PBO is more flexible with respect to adapting the configuration, which can be necessary due to geometrical constraints \cite{medani2012,medani2015}. Further  development of these interpolation  techniques is an interesting future goal. 

With regard to the computational workload, all the investigated source modeling methods are essentially similar, since by far the greatest work needed with each approach is to compute the transfer matrix ${\bf T}$ (Section \ref{transfer_matrix}) corresponding to the nodal basis, although the model-dependent load vector ${\bf f}$ can be formed with much less of effort, see e.g. \cite{bauer2015}. Calculating the load vector directly as a linear combination of the nodal basis functions is also computationally fast, e.g., in comparison to the indirect subtraction approach  \cite{CHW:Dre2009,CHW:Lew2009b,CHW:Awa97,CHW:Sch2002,brazier1961}, where higher-order numerical integration techniques are needed to estimate a correction term. With respect to the other implementation aspects, the present vector approach does require slightly more extensive data structuring compared to the monopolar methods, since the lists of mesh faces and edges are needed in addition to those of nodes and tetrahedra. 

Finally, this study works as an important proof-of-concept for the future  development of potential FEM based EEG forward modeling techniques, for example, the discontinuous Galerkin (DG) and the Mixed-FEM \cite{vorwerk2016a,vorwerk2016b,engwer2015} which are advantageous for modeling current fields and can   necessitate vector field basis functions to be used. Other future directions will include modeling of finitely supported primary currents and inversion of neural activity utilizing the present divergence conforming $H$(div) approach. One can, for example, study the recovery of a complete vector field instead of a set of individual dipolar sources. Applying and evaluating the newly developed FI and EW FEM source models and interpolation strategies within a realistic head geometry are particularly attractive and also necessary future goals due to the many simplifications of the  Stok model. For example, focality with respect to the conductivity jumps needs to be elucidated. Additionally, improving the PBO and MPO interpolation techniques as well as the electric potential and primary current field function bases, e.g., via higher-order bubble functions, would be interesting topics to be studied in the future.

\section{Appendix}
\label{appendix}

\subsection{Linear Vector Basis Functions}

\begin{figure}[h]
\begin{center} 
\begin{minipage}{3.7cm}
\begin{center}
    \includegraphics[height=4.1cm]{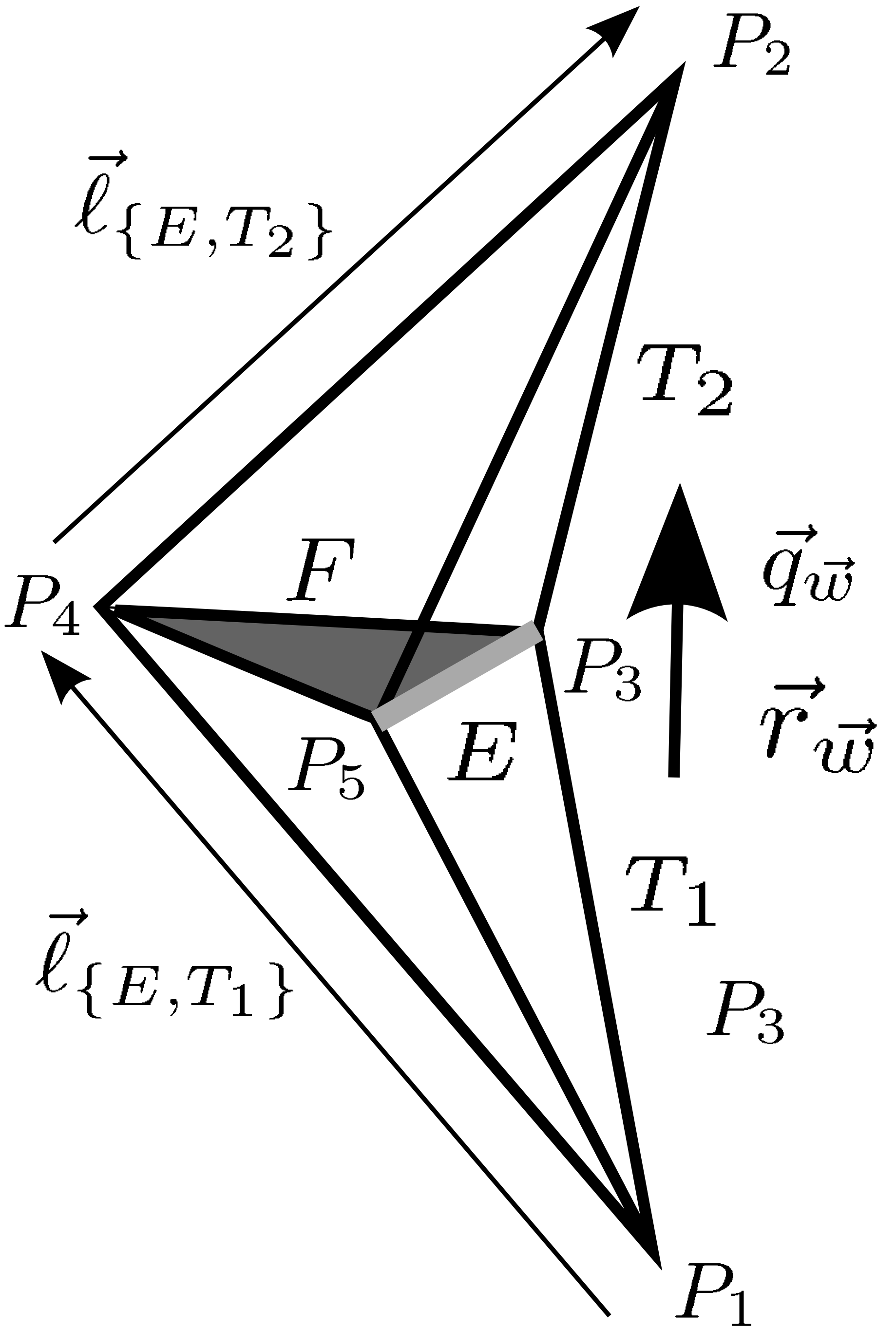} 
\end{center}
\end{minipage}  
\begin{minipage}{3.7cm}
\begin{center}
\includegraphics[height=3.2cm]{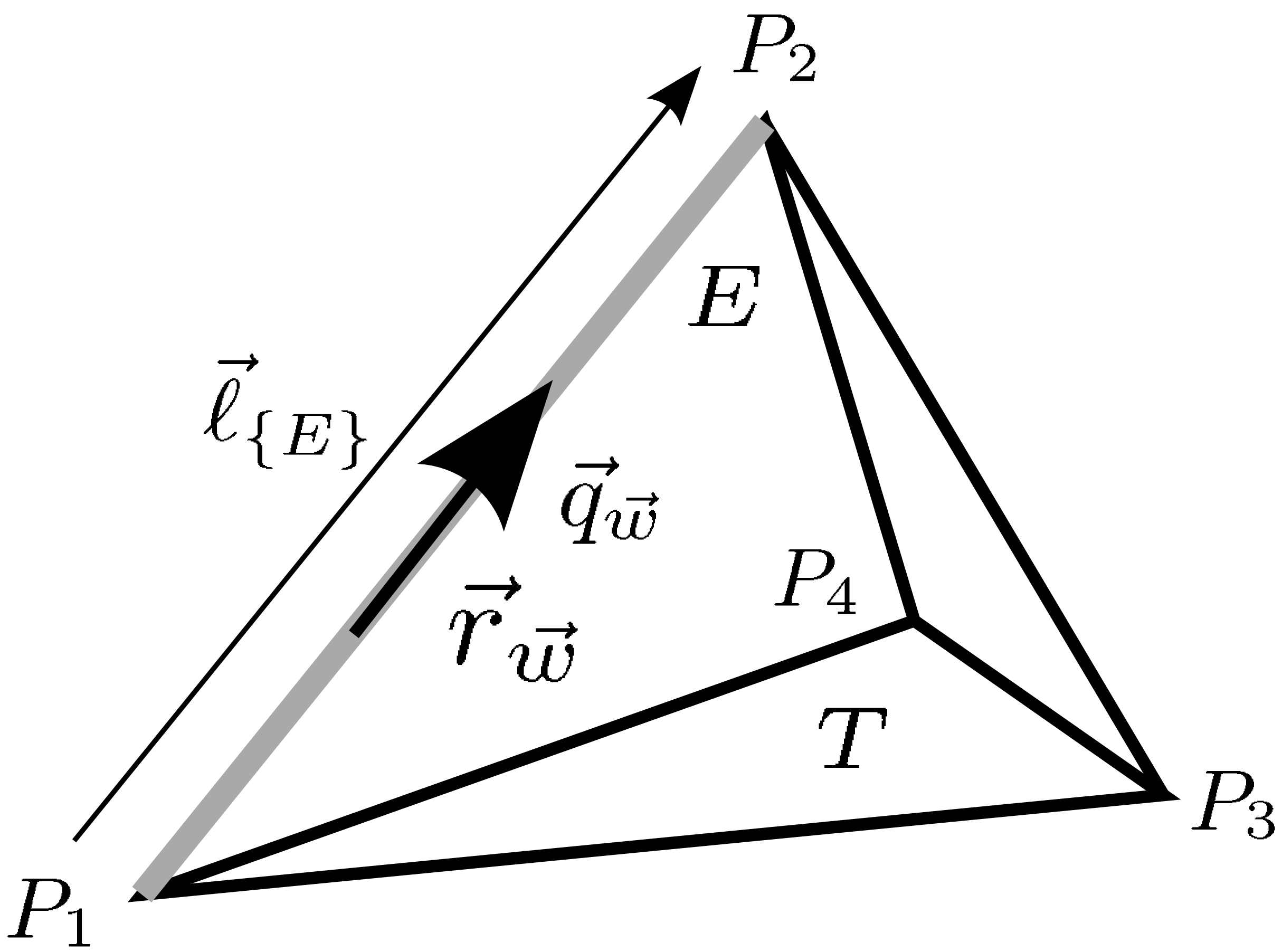}
\end{center}
\end{minipage} \\ \vskip0.3cm
\begin{minipage}{3.7cm}
\begin{center}
    \includegraphics[width=2.8cm]{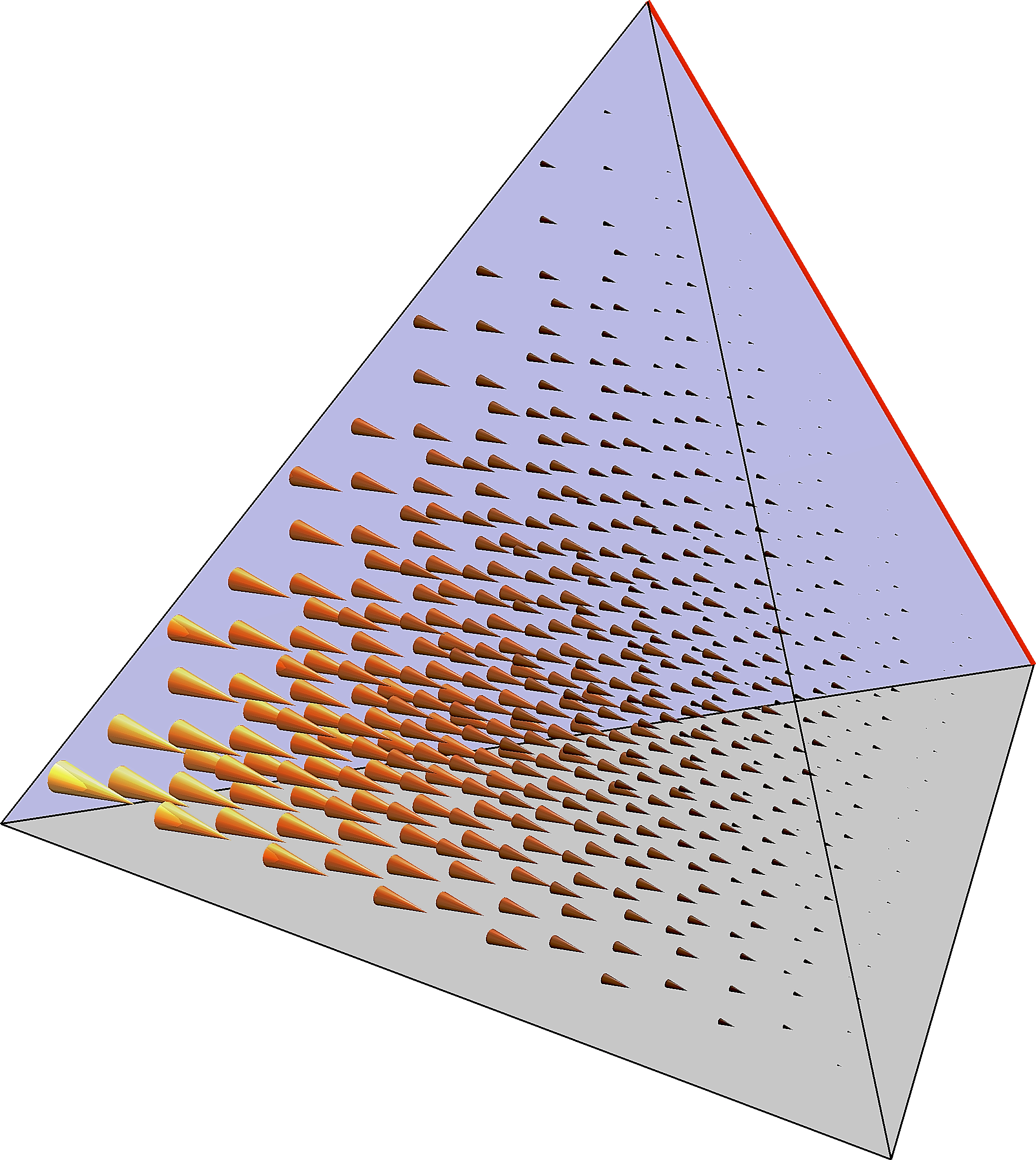} \\ \vskip0.2cm
Linear 
\end{center}
\end{minipage}  
\begin{minipage}{3.7cm}
\begin{center}
\includegraphics[width=2.78cm]{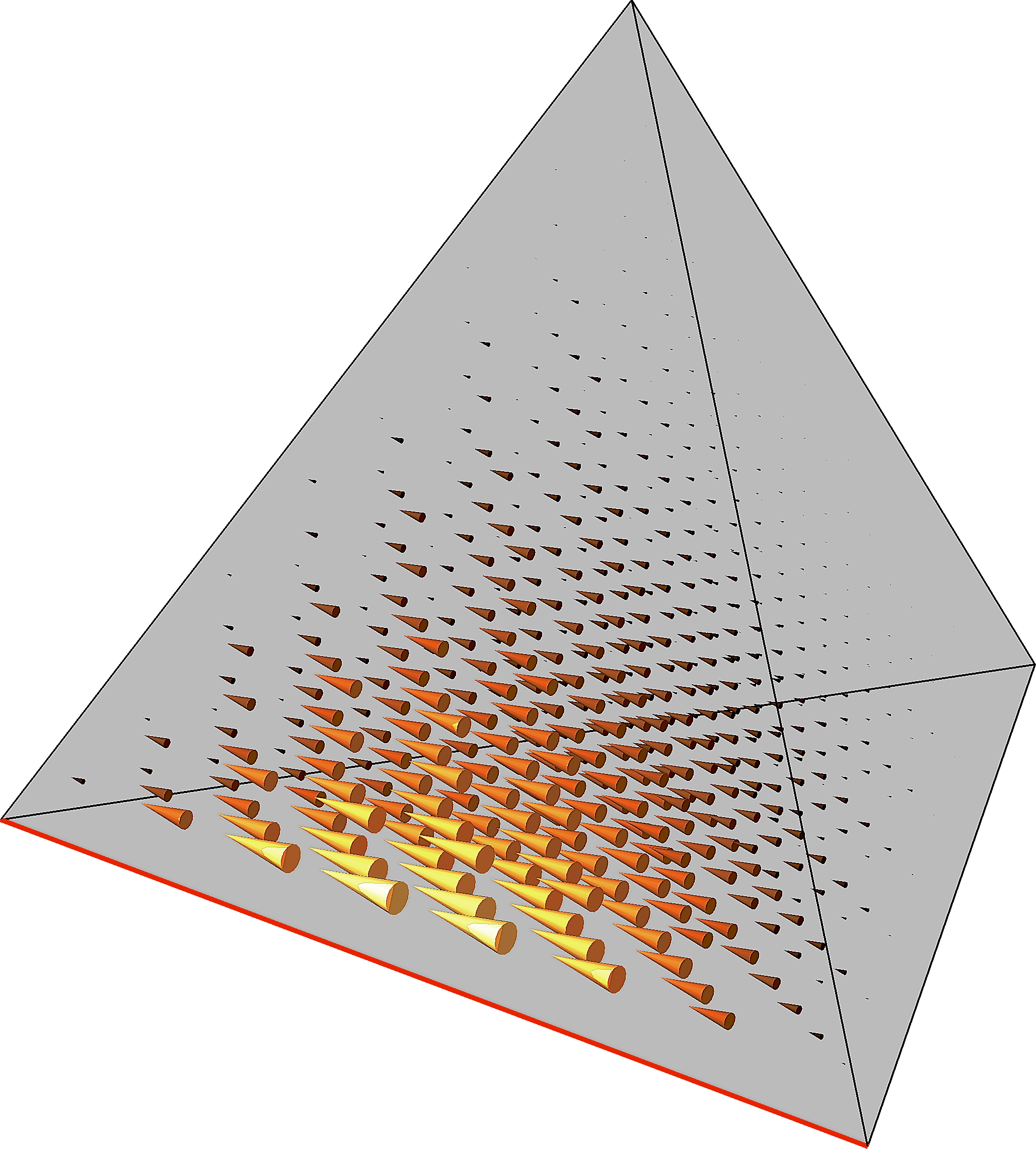}\\ \vskip0.2cm
Quadratic
\end{center}
\end{minipage}
\end{center}
    \caption{Linear edge-based face functions (left) and quadratic edge-based interior functions (right). Top row includes a schematic visualization of the support and the features essential for the definition.  Edge $E$ is visualized by the bold (light grey) line. Face $F$ has been darkened, and  edge vectors $\vec{\ell}_{E,T_1}$, $\vec{\ell}_{E,T_2}$  and $\vec{\ell}_{E}$ are shown by the thin arrows. The resulting  dipolar source  characterized by $q_{\vec{w}}$ and $\vec{r}_{\vec{w}}$ is indicated by the bold arrow. Bottom row shows the corresponding vector field, when limited to a single tetrahedron $T$. }
    \label{fig:nedelec} \label{nedelec}
\end{figure}

The piecewise linear subspace of $H(\hbox{div})$ for a tetrahedral mesh is spanned by linear N\'ed\'elec's edge-based face functions \cite{monk2003,ainsworth2003} (Figure \ref{nedelec}). Each of these  is supported in two tetrahedra $T_1$ and $T_2$ that share the face $F$ (Figure \ref{nedelec}).  A basis function within a single tetrahedron $T$ is of the form 
\begin{equation}
\vec{w}_{ \{ E, F, T \}} = {c_{\{ E, F \}}}   \; \psi_{\{ F, T \} } \; \frac{\vec{\ell}_{\{ E, T \} }}{V_T}
\end{equation} 
where face $F$, edge $E$ and edge vector $\vec{\ell}_{E,T}$ are as given in Figure \ref{nedelec},  $V_T$ is the volume of $T$, and  $\psi_{\{ F, T \} }$ is the linear nodal basis function in $T$ associated with the node opposite to $F$. 
Defining 
\begin{equation} c_{\{ E,F\}}  =  \frac{4}{\| \vec{\ell}_{\{E,T_1\}}  + \vec{\ell}_{\{E,T_2\}} \|} = \frac{4}{\| \vec{r}_{P_2}  - \vec{r}_{P_1} \|}, 
\end{equation}  it follows that the dipolar moment  is a unit vector, as shown by 
{\setlength\arraycolsep{2 pt} 
\begin{eqnarray}
\vec{q}_{\vec{w}}  & = &  \int_\Omega \vec{w} \; dV = \int_{T_1} \vec{w}_{ \{  E, F, T_1 \}} \; dV +\int_{T_2} \vec{w}_{ \{  E, F, T_2 \}} \; dV \nonumber\\ & = & \; \; \; \; c_{\{ E,F \}} \;   \frac{\vec{\ell}_{\{E,T_1\}}}{V_{T_1}} \; \int_{T_1} \psi_{\{ F, T_1 \} } \; dV \nonumber\\ & & + \;  c_{\{ E,F \}}  \;   \frac{\vec{\ell}_{\{E,T_2\}}}{V_{T_2}} \; \int_{T_2} \psi_{\{  F, T_2 \} } \; dV \nonumber\\ & = & \; \; \; \; c_{\{ E,F \}} \; \frac{\vec{\ell}_{\{E,T_1\}}}{V_{T_1}} \; \frac{V_{T_1}}{4} + \;  c_{\{ E,F \}} \; \frac{\vec{\ell}_{\{E,T_2\}}}{V_{T_2}} \; \frac{V_{T_2}}{4}  \nonumber\\ 
& = & \frac{\vec{\ell}_{\{E,T_1\}}  + \vec{\ell}_{\{E,T_2\}}}{\| \vec{\ell}_{\{E,T_1\}}  + \vec{\ell}_{\{E,T_2\}} \|} =  \frac{\vec{r}_{P_2}  - \vec{r}_{P_1}}{\| \vec{r}_{P_2}  - \vec{r}_{P_1} \|}.
\end{eqnarray}} 
Notice that any linear nodal basis function $\psi$ integrated over $T$ equals $V_T/4$. 
Furthermore, it holds that $\nabla \psi_{\{ F, T_2 \} }\cdot {\vec{\ell}_{\{E, T_2\}}}= -\nabla \psi_{\{ F, T_1 \} }\cdot {\vec{\ell}_{\{E, T_1 \}}} = 1$, since the linear function $\psi_{\{  F, T \} }$ increases from zero to one on a path corresponding to (positive or negative) vector ${\vec{\ell}_{\{E, T\}}}$. Consequently, one has 
{\setlength\arraycolsep{2 pt} 
\begin{eqnarray}
 G_{\psi,\vec{w}} & = & - \int_\Omega (\nabla \cdot \vec{w}) \; \psi \; dV \nonumber\\  
& = & - c_{\{  E, F \}}  \;\nabla \psi_{\{  F, T_2 \} }\cdot \frac{\vec{\ell}_{\{E, T_2\}}}{V_{T_2}} \; \int_{T_2}  \; \psi \; dV \nonumber\\  
&  & - c_{\{ E, F \}}  \; {\nabla \psi_{\{ F, T_1 \} } \cdot  \frac{\vec{\ell}_{\{E, T_1\}}}{V_{T_1}}  } \; \int_{T_1} \;  \psi \; dV \nonumber\\ 
&=&    \frac{s_{\{ \psi, P_2 \}} \! - \! s_{\{ \psi, P_1\} }}{ \| \vec{\ell}_{\{E,T_2\}}  + \vec{\ell}_{\{E,T_1\}} \|} = \frac{s_{\{ \psi, P_2 \}} \! - \! s_{\{ \psi, P_1\} }}{\| \vec{r}_{P_2}  - \vec{r}_{P_1} \|}. 
\end{eqnarray}}

\subsection{Quadratic Vector Basis Functions}

Edge-based interior functions complement the piecewise linear subspace to a quadratic one  \cite{ainsworth2003}. Each one of these is supported on the set of $\mathfrak{n}$ tetrahedra $T_1, T_2, \ldots, T_{\mathfrak{n}}$ sharing edge $E$ (Figures \ref{fig:nedelec}). Restricted to a single tetrahedron, a basis function is  of the form 
\begin{equation}
\vec{w}_{ \{ E,  T \}} = {c_{\{ E \}}}   \; \psi_{\{ E, T, P_1 \} } \psi_{\{ E, T, P_2 \} }  \; \frac{\vec{\ell}_{\{ E \} }}{ V_T}, 
\end{equation}
where $P_1$ and $P_2$ are the end points of edge $E$ as shown in Figure \ref{fig:nedelec}, and $\psi_{\{ E, T, P \} } $ is the nodal basis function associated with point $P$.   
Choosing \begin{equation}
c_{\{ E\}}  =  \frac{20}{ \mathfrak{n} \, \| \vec{\ell}_{\{E\}}  \| } =  \frac{20}{\mathfrak{n} \, \| \vec{r}_{P_2}  - \vec{r}_{P_1} \| } \end{equation}  yields a unit-length  dipolar moment  given by 
{\setlength\arraycolsep{2 pt} 
\begin{eqnarray}  \vec{q}_{\vec{w}}  & =& \int_\Omega \vec{w} \; dV =  \sum_{\ell = 1}^\mathfrak{n} \, \int_{T_\ell} \vec{w}_{ \{  E, T_\ell \}} \; dV
  \nonumber\\ & = & \; \; \; \; c_{\{ E \}} \;   \sum_{\ell = 1}^{\mathfrak{n}} \frac{\vec{\ell}_{\{E\}}}{V_{T_\ell}} \; \int_{T_\ell} \psi_{\{ E,P_1 \} } \psi_{\{ E, P_2 \} }  \; dV \nonumber\\ & = & \; \; \; \; c_{\{ E \}} \; \frac{\mathfrak{n} \, \vec{\ell}_{\{E\}}}{V_{T}} \; \frac{V_{T}}{20} =  \frac{\vec{\ell}_{\{E\}} }{\| \vec{\ell}_{\{E\}} \|} = \frac{\vec{r}_{P_2}  - \vec{r}_{P_1}}{\| \vec{r}_{P_2}  - \vec{r}_{P_1} \|}.
\end{eqnarray}} 
Furthermore,  it follows that
{\setlength\arraycolsep{2 pt} 
\begin{eqnarray}
 G_{\psi,\vec{w}} & = & - \int_\Omega (\nabla \cdot \vec{w}) \; \psi \; dV \nonumber\\  
& = & - c_{\{  E\}}  \;\nabla \psi_{\{ E,  P_1 \} } \! \cdot \! \frac{\vec{\ell}_{\{E, T\}}}{V_{T}} \; \! \sum_{\ell = 1}^\mathfrak{n} \int_{T_\ell}  \! \! \! \psi_{\{ E, P_2 \} } \psi_{\{ E,  P_j \} } \; dV \nonumber\\  
&  & - c_{\{ E \}}  \; {\nabla \psi_{\{ E,  P_2 \} } \! \cdot \!  \frac{\vec{\ell}_{\{E, T\}}}{V_{T}}  } \; \! \sum_{\ell = 1}^\mathfrak{n} \int_{T_\ell}  \! \! \!  \psi_{\{ E, P_2 \} } \psi_{\{ E, P_j \} } \; dV \nonumber\\ 
&=&    \frac{s_{\{ \psi,  P_2\}}  - \! s_{\{ \psi, P_1\} }}{ \| \vec{\ell}_{\{E\}} \|} = \frac{s_{\{ \psi,  P_2\}}  - \! s_{\{ \psi, P_1\} }}{\| \vec{r}_{P_2}  - \vec{r}_{P_1} \|}, 
\end{eqnarray}}
where the fact that $\int_{T_1}  \; \psi_{\{ E,  P_i \} } \psi_{\{ E, P_ j \} } \; dV = V_T/10$, if $i=j$, and  $\int_{T_1}  \; \psi_{\{ E,  P_i \} } \psi_{\{ E, P_ j \} } \; dV = V_T/20$ otherwise, has been used.

\section*{Acknowlegment}

SP was supported by the Academy of Finland (project 257288). JV and CW were supported by the Priority Program 1665 of the Deutsche Forschungsgemeinschaft (DFG) (project WO1425/5-1) and by EU project ChildBrain (Marie Curie Innovative Training Networks, grant agreement no. 641652).

\section*{References}

\bibliographystyle{jphysicsB_withTitles}
\bibliography{bibliography,chw,pursiainen}

\end{document}